\documentclass[12pt,twoside]{article}
\usepackage{amssymb,maxitpb}
\def\fHH{\|\f\|^2_2}
\def\fTH{\|\f\|^1_2}
\def\tN{\tilde\nabla}

\def\SF{\Sigma\Lambda}
\def\SFtop{\Sigma\Lambda_{{\rm top}}}
\def\td{\tilde d}
\def\tdtop{\tilde d^{{\rm top}}}
\def\eg{\varepsilon(\gamma)}
\def\ig{\iota(\gamma)}
\def\tD{\tilde\delta}

\def\YYb{\underline{\bf Y}}

\def\YZb{\underline{\bf Z}}

\def\gSS{{\cal S}_{{\bf\S}}}
\def\gTT{{\cal S}_{{\bf T}}}
\def\gYY{{\cal S}_{{\bf Y}}}
\def\gZZ{{\cal S}_{{\bf Z}}}

\def\gYYZ{{\cal G}_{\underline{{\bf Z}}}}
\def\gZYZ{{\cal H}_{\underline{{\bf Z}}}}
\def\gYYY{{\cal G}_{\underline{{\bf Y}}}}

\font\bldfc=cmmib10
\def\boldface#1{\mbox{\bldfc#1}}
\def\bold{\boldface}

\newcommand{\XA}%
{\truex{300}%
\begin{picture}(0,0)%
\put(0,0){\circle{\value{x}}}%
\end{picture}}

\newcommand{\XB}%
{\truex{500}%
\begin{picture}(0,0)%
\put(0,0){\circle*{\value{x}}}%
\end{picture}}

\newcommand{\MV}[3]{\movevertex(#1,#2){#3}}

\def\CN{\cross{}{\bold{\nar[60]}}}

\def\CS{\cross{}{\bold{\sar[60]}}}
\def\CSW{\cross{}{\bold{\swar[60]}}}
\def\CSE{\cross{}{\bold{\sear[60]}}}

\def\CE{\cross{}{\bold{\ear[60]}}}

\newcommand{\nnn}[1]{(\ref{#1})}
\newcommand{\text}[1]{\mbox{#1}}
\newcommand{\dfrac}[2]{{\displaystyle{\frac{#1}{#2}}}}
\newcommand{\tfrac}[2]{{\textstyle{\frac{#1}{#2}}}}

\newtheorem{theorem}{Theorem}
\newtheorem{lemma}[theorem]{Lemma}
\newtheorem{proposition}[theorem]{Proposition}

\newtheorem{corollary}[theorem]{Corollary}

\newtheorem{remark}[theorem]{Remark}

\newenvironment{proof}[1]{\begin{trivlist} \item[] {\em #1}. }%
{\hfill $\square$ \end{trivlist}}

\begin{document}
\def\thefootnote{}
\footnotetext{Both authors are partially supported by NATO Collaborative 
Research Grant 960129.  TB supported by NSF Grant INT-9722779.}
\def\thefootnote{\dagger}

\let\chap=\S

\let\tilde=\widetilde
\let\implies=\Longrightarrow

\let\a=\alpha
\let\b=\beta
\let\d=\delta
\let\D=\Delta
\let\e=\varepsilon
\let\f=\varphi
\let\F=\Phi
\let\g=\gamma
\let\G=\Gamma
\let\h=\ell
\let\i=\iota
\let\k=\kappa
\let\l=\lambda
\let\L=\Lambda
\let\nd=\nabla
\let\N=\nabla
\let\m=\mu
\let\r=\rho
\let\s=\sigma
\let\S=\Sigma
\let\t=\tau
\let\w=\omega
\let\W=\Omega
\let\x=\xi
\let\y=\psi
\let\z=\zeta

\def\ath{a^{\underline{{\rm th}}}}
\let\bop=\bigoplus
\def\cason{{\rm Cas}_{{\mathfrak{so}}(n)}}
\def\ci{C^\infty}
\def\circe{{\scriptstyle\circ}}
\def\congn{\cong_{{\rm Spin}(n)}}
\def\Cop{C^{{\rm op}}}
\let\da=\downarrow
\def\dc{\mbox{$\nabla\!\!\!\! /\;$}}
\def\rso{{\cal S}^0}

\def\injsptocottw{I_{{\bf\S}}}
\def\injtwtocottw{I_{{\bf T}}}

\def\df{{\displaystyle{\frac12}}}
\def\dv{{\rm div}}
\def\dvol{d{\rm vol}}
\def\eye{\sqrt{-1}}
\def\gll{G_{\l\l}}
\let\implies=\Rightarrow
\def\ip#1#2{\langle#1,#2\rangle}
\let\lra=\leftrightarrow
\def\nl{\hfill\break}
\let\os=\oplus
\let\ot=\otimes
\def\Rop{R^{{\rm op}}}
\def\spn{{\rm Spin}(n)}
\def\spN{{\rm Spin}(n+1)}
\def\swu{G_u^*G_u}
\def\tf{{\textstyle{\frac12}}}
\let\ua=\uparrow

\def\bbC{{\mathbb{C}}}
\def\bbD{{\mathbb{D}}}
\def\bbN{{\mathbb{N}}}
\def\bbR{{\mathbb{R}}}
\def\bbT{{\mathbb{T}}}
\def\bbU{{\mathbb{U}}}
\def\bbV{{\mathbb{V}}}
\def\bbW{{\mathbb{W}}}
\def\bbZ{{\mathbb{Z}}}

\def\bT{{\cal T}}

\def\cD{{\cal D}}
\def\cE{{\cal E}}
\def\cF{{\cal F}}
\def\cG{{\cal G}}
\def\cH{{\cal H}}
\def\cJ{{\cal J}}
\def\cL{{\cal L}}
\def\cM{{\cal M}}
\def\cN{{\cal N}}
\def\cP{{\cal P}}
\def\cR{{\cal R}}
\def\cS{{\cal S}}
\def\cT{{\cal T}}
\def\cW{{\cal W}}
\def\cY{{\cal Y}}

\def\dls{\cS_\l}

\def\gg{{\mathfrak{g}}}
\def\gso{{\mathfrak{so}}} 

\def\endo{{\rm End}}
\def\hom{{\rm Hom}}
\def\id{{\rm Id}}
\def\ord{{\rm ord}}
\def\orth{{\rm O}}
\def\pin{{\rm Pin}}
\def\proj{{\rm Proj}}
\def\sgn{{\rm sgn}}
\def\so{{\rm SO}}
\def\spam{{\rm span}}
\def\spin{{\rm Spin}}
\def\sym{{\rm Sym}}
\def\tc{\tilde c}
\def\tfs{{\rm TFS}}
\def\tr{{\rm tr}}
\def\Tr{{\rm Tr}}

\def\bdm{\begin{displaymath}}
\def\edm{\end{displaymath}}
\def\beq{\begin{equation}}
\def\eeq{\end{equation}}

\def\Sp{{\bf\S}}
\def\Tw{{\bf T}}

\def\Yb{{\bf Y}}
\def\Zb{{\bf Z}}
\def\pS{{\bf P}_{{\bf\S}}}
\def\pT{{\bf P}_{{\bf T}}}
\def\pY{{\bf P}_{{\bf Y}}}
\def\pZ{{\bf P}_{{\bf Z}}}
\def\agS{{\bf G}_{{\bf\S}}^*}
\def\gS{{\bf G}_{{\bf\S}}}
\def\gT{{\bf G}_{{\bf T}}}
\def\gY{{\bf G}_{{\bf Y}}}
\def\gZ{{\bf G}_{{\bf Z}}}
\def\GS{\gS^*\gS}
\def\GT{\gT^*\gT}
\def\GY{\gY^*\gY}
\def\GZ{\gZ^*\gZ}

\def\ET{{\bf E}_{{\bf T}}}

\def\xint{0}
\def\xfoo{1}
\def\xfur{2}
\def\xbun{3}
\def\xwei{4}
\def\xove{5}
\def\xspec{6}
\def\xmix{7}

\begin{center}
{\large Bochner-Weitzenb\"ock formulas associated with 
the Rarita-Schwinger operator}\\[15pt]
{\large Thomas Branson and Oussama Hijazi
}
\end{center}
\tableofcontents 
\newpage
\section{Introduction}

In this paper, we establish basic material for future investigations
of the analysis and geometry of the twistor bundle, and of differential
operators with the twistor bundle as source and/or target, especially
the {\em Rarita-Schwinger operator}, a first order differential operator
taking twistors to twistors.  Some 
of the material that we shall present
generalizes to arbitrary irreducible tensor-spinor
bundles \cite{tbsw}.  In addition, some material which does not have
a clear generalization of this breadth should nevertheless extend to 
statements about spinor-forms (see \cite{tbadv}), or about bundles
contained in the tensor product of the spinors with the trace-free
symmetric tensors.  There are some nice complementary 
results in more analytic directions for flat structures; see 
for example \cite{soubook} and \cite{bssv}.
One direct inspiration for our investigations
is the success of the spinor program \cite{BFGK,BHMM}, and we have been guided
by a desire to obtain analogues of the most important results of this
field.  

Some of the results we state here are undoubtedly not the most refined
or extensive possible.  However, the relevant identities and
decompositions do not seem to be in general circulation.  Given this, 
it seems timely to
put some of this material into print, together with sufficient
concrete results to indicate the motivation and effectiveness of the
method.  One of our main themes is that to ``work on twistors and the
Rarita-Schwinger operator,''  one
needs to also consider several other related bundles and operators.

\section{Familiar vector bundles and first-order differential
operators}\setcounter{equation}0
Let $(M,g,E,\g)$ be an $n$-dimensional Riemannian spin manifold.  
That is, we have a Riemannian manifold $(M,g)$ which admits
spin structure, and thus has a volume $n$-form $E$, a spinor bundle $\Sp$,
and a {\em fundamental tensor-spinor} 
$\g$; this is a smooth section of the bundle 
$TM\ot\endo(\Sp)$ with
\bdm
\g^\a\g^\b+\g^\b\g^\a=-2g^{\a\b}\qquad\mbox{and}\qquad\N\g=0.
\edm
The connection $\N$
 is the natural extension of the Levi-Civita connection on $TM$
to tensor-spinors of arbitrary type. Here and below, we use an abstract 
tensor index notation, but do not
write spinor indices explicitly.  ``Abstract'' is meant in the sense
of Penrose (\cite{pen}, \chap2): the indices do not refer to a choice
of local frame, but rather are placeholders; indicating, among other
things, how to compute the expression locally should choices be made.
The allowable manipulations may then be described by a finite number
of axioms.  The dimension, but not the
signature of the metric, is detectable via such manipulations.  As
usual in tensor calculus, an expression like $\N_\a\w_\b$ denotes
$(\N\w)_{\a\b}\,$.  An index which appear twice in a term, once up and
once down, indicates a contraction, and indices may be lowered and
raised using the metric tensor and its inverse.  

Given a vector bundle ${\bf V}$, we denote by $\G({\bf V})$ its smooth 
section space.
The {\em Dirac operator} is, up to normalization, the operator
\bdm
\begin{array}{rl}
\dc:\G(\Sp)\to&\G(\Sp), \\
\y\mapsto&\g^\a\N_\a\y.
\end{array}
\edm   
Let $\Tw$ be the {\em twistor bundle}; that is, the 
subbundle of
$T^*M\otimes\Sp$ determined by the (pointwise) equation 
\bdm\g^\a\f_\a=0\,.
\edm
The {\em twistor operator} is the operator
\beq\label{twistop}
\begin{array}{rl}
\bT:\G(\Sp)\to&\G(\Tw), \\
\y\mapsto&\N_\a\y+\dfrac1n\g_\a\;\dc\y.
\end{array}
\eeq
The formal adjoint of the twistor operator is
\beq\label{adjtw}
\begin{array}{rl}
\bT^*:\G(\Tw)\to&\G(\Sp), \\
\f\mapsto&-\N^\a\f_\a\,,
\end{array}
\eeq
as one sees via the calculation
\bdm
\begin{array}{rl}
\left\langle\left(\N_\a+\dfrac1n\g_\a\;\dc\right)\y,\f^\a\right\rangle
&=-\left\langle\y,\N_\a\f^\a+\dfrac1n\;\dc(\g_\a\f^\a)\right\rangle
+\mbox{(exact divergence)} \\
&=-\left\langle\y,\N^\a\f_\a\right\rangle+\mbox{(exact divergence)} \\
&=\left\langle\y,\N^*\f\right\rangle+\mbox{(exact divergence)}.
\end{array}
\edm
Here we have used the covariant constancy of the spin metric
$\ip{\cdot}{\cdot}$, the skew-adjointness of $\g^\a$ (and the consequent
formal self-adjointness of $\dc$), and the fact
that $\g^\a\f_\a=0$ for a section of $\Tw$.  
By ``exact divergence'', we mean an expression of the form
$\N^\a\w_\a\,$, where $\w\in\G(T^*M)$.
We shall often make calculations
like this, without explicitly noting all the steps.

The operator $\bT$ may be described as $P\circe\N$, where $\N$ is the
covariant derivative $\G(\Sp)\to\G(T^*M\ot\Sp)$, and $P$ is the
orthogonal projection of $T^*M\ot\Sp$ onto $\Tw$.  Since $\Tw$ is a
$\spn$-subbundle of $T^*M\ot\Sp$, the projection $P$ is
$\spn$-equivariant.  Given this, the formula \nnn{adjtw} is not
surprising, and is an example of a more general phenomenon: since
orthogonal projections are self-adjoint,
$(P\circe\N)^*=\N^*\circe P=\N^*$ on $\G(\Tw)$.

The $\spn$-bundle complementary to $\Tw$ is the image of $\Sp$ under
the injection $I:\y\mapsto\g_\a\y$.  This map is clearly $\spn$-equivariant.  
Injectivity is
guaranteed by the calculations
\bdm
I^*\,\f = - \g_{\a} \,\f^\a \,,\qquad I^*\,I = n \,\id_{\Sp} \,, 
\edm
which show that $n^{-1/2}I$ is an {\em isometric} injection. 

In view of \nnn{twistop},
the corresponding decomposition
of $\N\y$ is
\bdm
\N\y=\bT\y-\dfrac1nI\;\dc\y.
\edm

The {\em Rarita-Schwinger} operator (see \nnn{ident}) on $\G(\Tw)$ is, up 
to normalization, the operator
\beq\label{rar1}
\begin{array}{rl}
\rso:\G(\Tw)\to&\G(\Tw), \\
\f\mapsto&\g^\l\N_\l\f_\a-\dfrac2n\g_\a\N^\l\f_\l\,.
\end{array}
\eeq
The operator $\rso$, like $\dc$, is formally 
self-adjoint.  It may be
described (and will be below) as the orthogonal projection of the operator
$\g^\l\N_\l$ on $\G(\Tw)$ to the (unique) subbundle $\bbW$ 
of $T^*M\ot\Tw$ which
is isomorphic to $\Tw$, followed by a bundle isomorphism $\bbW\to\Tw$.

\section{Further relevant bundles and operators}\setcounter{equation}0
Let $\tfs^2$ be the bundle of trace-free symmetric two-tensors,
and let
$\Zb$ be the subbundle of $\tfs^2\ot\Sp$ determined by the pointwise
condition
\beq\label{annih}
\g^\b\F_{\a\b}=0.
\eeq
With \nnn{annih} in place, 
the trace-free condition is actually redundant,
since by the Clifford relations, 
$0=\g^\b\g^\a\F_{\a\b}=-g^{\a\b}\F_{\a\b}\,$.
Note that \nnn{annih} requires $\F$ to also be a section of $T^*M\ot\Tw$,
so that 
$$ 
\Zb= ( \tfs^2\ot\Sp )\cap ( T^*M\ot\Tw ).
$$
Similarly, let 
$\Yb$ be the subbundle of $\L^2\ot\Sp$ determined by the condition
\nnn{annih}; that is, 
$$
\Yb=( \L^2\ot\Sp )\cap ( T^*M\ot\Tw ).
$$
Two more subbundles of $T^*M\ot\Tw$ may be defined by injecting 
$\Sp$ into $T^*M\ot\Tw$ using the map
\bdm
\y\mapsto(
\injsptocottw
\y)_{\a\b}=
\left\{\g_\a\g_\b+\dfrac{n-2}{n}\g_\b\g_\a\right\}\y,
\edm
and by injecting $\Tw$ into $T^*M\ot\Tw$ using the map
\bdm
\f_\a\mapsto(
\injtwtocottw
\f)_{\a\b}=\g_\a\f_\b-\dfrac2n\g_\b\f_\a\,.
\edm
The maps $\injsptocottw$ and $\injtwtocottw$ 
are clearly $\spn$-equivariant.  Injectivity 
is
guaranteed by the calculations
\beq\label{isoinj}
\begin{array}{rlrl}
\injsptocottw^*\,\F&=-2\F^\a{}_\a\,,\qquad & \injsptocottw^*\injsptocottw
&=4(n-1)\,\id_{\Sp}\,, 
\\
(\injtwtocottw^*\F)_\a&=-\g^\b\F_{\b\a}\,,\qquad &
\injtwtocottw^*\injtwtocottw
&=\dfrac{(n+2)(n-2)}n\,
\id_{\Tw}\,,
\end{array}
\eeq
which also show that $\{4(n-1)\}^{-1/2}\injsptocottw$ 
and $\{(n+2)(n-2)/n\}^{-1/2}\injtwtocottw$
are {\em isometric} injections.

For use in some of the following formulas, we introduce the {\em antisymmetric
Clifford symbols}
$$
\g_{\a\b}:=\dfrac12(\g_\a\g_\b-\g_\b\g_\a),
$$
so that
$$
\g_\a\g_\b = \g_{\a\b} - g_{\a\b}.
$$
The four subbundles of $T^*M\ot\Tw$ given above are clearly orthogonal.
Moreover, the maps $\pS\,$, $\pT\,$,
$\pY\,$, $\pZ\,$ given by
\bdm
\begin{array}{rl}
(\pS\F)_{\a\b}&=-\dfrac1{2(n-1)}\left\{\g_\a\g_\b+\dfrac{n-2}n\g_\b\g_\a
\right\}\F^\l{}_\l 
 \\
 &=\dfrac1{n}\left\{g_{\a\b}-\dfrac1{n-1}\g_{\a\b}\right\}\F^\l{}_\l\,,
\end{array}
\edm
\bdm
\begin{array}{rl}
(\pT\F)_{\a\b}&=\dfrac1{(n+2)(n-2)}\Big\{-n\g_\a\g^\l\F_{\l\b}
+2\g_\b\g^\l\F_{\l\a}+2\g_\a\g_\b\F^\l{}_\l\\\
&\qquad\qquad\qquad\qquad-\dfrac4{n}\g_\b\g_\a\F^\l{}_\l\Big\} 
 \\
 &=\dfrac1{(n+2)(n-2)}\Big\{-n\g_\a{}^\l\F_{\l\b}+2\g_\b{}^\l\F_{\l\a}
 +n\F_{\a\b}-2\F_{\b\a}\Big\} \\
 &\qquad+\left(\dfrac2{n(n-2)}\g_{\a\b}
 -\dfrac2{n(n+2)}g_{\a\b}\right)\F^\l{}_\l\,,
\end{array}
\edm
\bdm
\begin{array}{rl}
(\pY\F)_{\a\b}&=\dfrac12\Big\{\F_{\a\b}-\F_{\b\a}\Big\}
+\dfrac1{2(n-2)}\Big\{\g_\a\g^\l\F_{\l\b}-\g_\b\g^\l\F_{\l\a}\Big\} \\
&\qquad-\dfrac1{2(n-1)(n-2)}\big\{\g_\a\g_\b-\g_\b\g_\a\big\}\F^\l{}_\l 
 \\
 &=\dfrac{n-3}{2(n-2)}\Big\{\F_{\a\b}-\F_{\b\a}\Big\}
 +\dfrac1{2(n-2)}\Big\{\g_\a{}^\l\F_{\l\b}-\g_\b{}^\l\F_{\l\a}\Big\} \\
 &\qquad-\dfrac1{(n-1)(n-2)}\g_{\a\b}\F^\l{}_\l\,,
\end{array}
\edm
\bdm
\begin{array}{rl}
(\pZ\F)_{\a\b}&=\dfrac12\Big\{\F_{\a\b}+\F_{\b\a}\Big\}
+\dfrac1{2(n+2)}\Big\{\g_\a\g^\l\F_{\l\b}+\g_\b\g^\l\F_{\l\a}
-2g_{\a\b}\F^\l{}_\l\Big\}
 \\
 &=\dfrac1{2(n+2)}\left\{(n+1)(\F_{\a\b}+\F_{\b\a})
 +\g_\a{}^\l\F_{\l\b}+\g_\b{}^\l\F_{\l\a}-2g_{\a\b}\F^\l{}_\l\right\}\,,
\end{array}
\edm
are complementary projections: one has
\bdm
\begin{array}{l}
\pS+\pT+\pY+\pZ=\id_{T^*M\ot\Tw}\,, \\
{\bf P}_u^2={\bf P}_u\,, \\
{\bf P}_u\,{\bf P}_v=0,\ u\ne v,
\end{array}
\edm
where $u,v$ run through the labels ${\bf\S}$, ${\bf T}$, ${\bf Y}$; and 
${\bf Z}$.
The projections $\pS\,$, $\pT\,$, $\pY\,$, and $\pZ$ are valued
in $\injsptocottw\Sp$, $\injtwtocottw\Tw$, $\Yb$, and $\Zb$ respectively.  
In particular,
$$
T^*M\ot\Tw=\injsptocottw\Sp\os\injtwtocottw\Tw\os\Yb\os\Zb\,.
$$
We define the first-order differential operators ${\bf G}_u$ by
applying the above projections to $\N\f$ for $\f\in\G(\Tw)$:
\beq\label{egS}
\begin{array}{rl}
(\gS\f)_{\a\b}&=-\dfrac1{2(n-1)}\left\{\g_\a\g_\b+\dfrac{n-2}{n}\g_\b\g_\a
\right\}\N^\l\f_\l \\ 
 &=\dfrac1{n}\left\{g_{\a\b}-\dfrac1{n-1}\g_{\a\b}\right\}\N^\l\f_\l\,,
\end{array}
\eeq
\beq\label{egT}
\begin{array}{l}
(\gT\f)_{\a\b}=\dfrac1{(n+2)(n-2)}\Big\{-n\g_\a\g^\l\N_\l\f_\b
+2\g_\b\g^\l\N_\l\f_\a+2\g_\a\g_\b\N^\l\f_\l\\
{}\qquad\qquad\qquad\qquad-\dfrac4{n}\g_\b\g_\a\N^\l\f_\l\Big\} \\
{}\qquad =
\dfrac1{(n+2)(n-2)}\left\{-n\g_\a{}^\l\N_\l\f_\b+2\g_\b{}^\l\N_\l\f_\a
 +n\N_\a\f_\b-2\N_\b\f_\a\right\} \\
{}\qquad\qquad+\left(\dfrac2{n(n-2)}\g_{\a\b}
 -\dfrac2{n(n+2)}g_{\a\b}\right)\N^\l\f_\l\,,
\end{array}
\eeq
\beq\label{egY}
\begin{array}{rl}
(\gY\f)_{\a\b}&=\dfrac12\left\{\N_\a\f_\b-\N_\b\f_\a\right\}
+\dfrac1{2(n-2)}\left\{\g_\a\g^\l\N_\l\f_\b-\g_\b\g^\l\N_\l\f_\a\right\} \\
&\qquad-\dfrac1{2(n-1)(n-2)}\left\{\g_\a\g_\b-\g_\b\g_\a\right\}\N^\l\f_\l \\
 &=\dfrac{n-3}{2(n-2)}\left\{\N_\a\f_\b-\N_\b\f_\a\right\}
 +\dfrac1{2(n-2)}\left\{\g_\a{}^\l\N_\l\f_\b-\g_\b{}^\l\N_\l\f_\a\right\} \\
 &\qquad-\dfrac1{(n-1)(n-2)}\g_{\a\b}\N^\l\f_\l\,,
\end{array}
\eeq
\beq\label{egZ}
\begin{array}{rl}
(\gZ\f)_{\a\b}&=\dfrac12\left\{\N_\a\f_\b+\N_\b\f_\a\right\}
+\dfrac1{2(n+2)}\Big\{\g_\a\g^\l\N_\l\f_\b\\
&\qquad\qquad\qquad\qquad\qquad+\g_\b\g^\l\N_\l\f_\a
-2g_{\a\b}\N^\l\f_\l\Big\} \\
 &=\dfrac1{2(n+2)}\Big\{(n+1)(\N_\a\f_\b+\N_\b\f_\a)
 +\g_\a{}^\l\N_\l\f_\b\\
&\qquad\qquad\qquad\qquad\qquad+\g_\b{}^\l\N_\l\f_\a-2g_{\a\b}\N^\l
\f_\l\Big\}\,.
\end{array}
\eeq
By the above remarks on formal adjoints, ${\bf G}_u^*{\bf G}_u
=\N^*{\bf G}_u$:
\bdm
\begin{array}{rl}
(\GS\f)_{\b}&=\dfrac1{2(n-1)}\Big\{\g_\a\g_\b+\dfrac{n-2}{n}\g_\b\g_\a
\Big\}\N^\a\N^\l\f_\l \\
 &=-\dfrac1{n}\left\{g_{\a\b}-\dfrac1{n-1}\g_{\a\b}\right\}\N^\a\N^\l\f_\l
\,,
\end{array}
\edm
\bdm
\begin{array}{rl}
(\GT\f)_{\b}&=\dfrac1{(n+2)(n-2)}\Big\{n\g_\a\g^\l\N^\a\N_\l\f_\b
-2\g_\b\g^\l\N^\a\N_\l\f_\a\\
&\qquad\qquad\qquad\qquad-2\g_\a\g_\b\N^\a\N^\l\f_\l
+\dfrac4{n}\g_\b\g_\a\N^\a\N^\l\f_\l\Big\} \\
 &=\dfrac1{(n+2)(n-2)}\Big\{n\g_\a{}^\l\N^\a\N_\l\f_\b
 -2\g_\b{}^\l\N^\a\N_\l\f_\a\\
 &\qquad\qquad\qquad\qquad  -n\N^\a\N_\a\f_\b+2\N^\a\N_\b\f_\a\Big\} \\
 &\qquad\qquad-\left(\dfrac2{n(n-2)}\g_{\a\b}
 -\dfrac2{n(n+2)}g_{\a\b}\right)\N^\a\N^\l\f_\l\,,
\end{array}
\edm
\bdm
\begin{array}{rl}
(\GY\f)_{\b}&=-\dfrac12\big\{\N^\a\N_\a\f_\b-\N^\a\N_\b\f_\a\big\}\\
&\qquad-\dfrac1{2(n-2)}\big\{
\g_\a\g^\l\N^\a\N_\l\f_\b-\g_\b\g^\l\N^\a\N_\l\f_\a\big\}\\
&\qquad+\dfrac1{2(n-1)(n-2)}\big\{\g_\a\g_\b-\g_\b\g_\a\big\}\N^\a\N^\l\f_\l 
\\ 
 &=-\dfrac{n-3}{2(n-2)}\big\{\N^\a\N_\a\f_\b-\N^\a\N_\b\f_\a\big\}\\
 &\qquad-\dfrac1{2(n-2)}\big\{\g_\a{}^\l\N^\a\N_\l\f_\b
 -\g_\b{}^\l\N^\a\N_\l\f_\a\big\} \\
 &\qquad+\dfrac1{(n-1)(n-2)}\g_{\a\b}\N^\a\N^\l\f_\l
\,,
\end{array}
\edm
\bdm
\begin{array}{rl}
(\GZ\f)_{\b}&=-\dfrac12\big\{\N^\a\N_\a\f_\b+\N^\a\N_\b\f_\a\big\}
-\dfrac1{2(n+2)}\big\{\g_\a\g^\l\N^\a\N_\l\f_\b\\
&\qquad+\g_\b\g^\l\N^\a\N_\l\f_\a
-2g_{\a\b}\N^\a\N^\l\f_\l\big\} \\
 &=-\dfrac1{2(n+2)}\Big\{(n+1)(\N^\a\N_\a\f_\b+\N^\a\N_\b\f_\a)\\
 &\qquad
 +\g_\a{}^\l\N^\a\N_\l\f_\b  + \g_\b{}^\l\N^\a\N_\l\f_\a
 -2g_{\a\b}\N^\a\N^\l\f_\l\Big\}\,.
\end{array}
\edm

The following is a consequence of the general elliptic
classification scheme of \cite{tbsw}, Theorem 4.10.  We also supply
an elementary proof below.

\begin{lemma}\label{strongell} 
The operator $\GT$ is strongly elliptic, in the sense that its leading
symbol is bounded below by a positive constant times the leading symbol
of $\N^*\N$.
\end{lemma}

\begin{proof}{Proof} One computes that if
\beq\label{Tws}
(T(\x)\f)_\b:=(n-1)\x^\a\x_\b\f_\a+\g_{\b\a}\x^\a\x^\l\f_\l\,,
\eeq
then $T(\x)^2=(n-1)|\x|^2T(\x)$, and
$$
\s_2(\GT)(\x)=-\dfrac1{(n+2)(n-2)}\left(-n|\x|^2+\dfrac4{n}T(\x)\right).
$$
As a result,
if 
$$
\ET:=-(n+2)^2\;\GT+\dfrac{2(n+2)(n^2-2n+2)}{n(n-2)}\;\N^*\N\,,
$$
then
$$
\s_2(\ET)(\x)\;\s_2(\GT)(\x)=|\x|^4.
$$
Thus $\s_2(\GT)(\x)$ is invertible for nonzero $\x$, so $\GT$
is elliptic.  Hence, since 
$\s_2(\GT)(\x)=\s_1(\gT)(\x)^*\s_1(\gT)(\x)$ is clearly positive
semidefinite, it is positive definite for $\x\ne 0$.
By equivariance, the eigenvalues of $\s_2(\GT)(\xi)$ have the form
$\m_i|\x|^2$, where the list of $\m_i$ is independent of $x\in M$
(and in fact independent of the manifold $M$).  By positive definiteness,
$0<\m:=\min\{\m_i\}$,
and we have
\beq\label{est}
\s_2(\GT) \ge \m \;\s_2(\N^*\N).
\eeq
\end{proof}

In fact, the proof of Lemma \ref{strongell} gives us more precise information:

\begin{corollary}\label{eigrk} For $a,b>0$, the operators 
$$
\GZ\ \ {\rm and}\ \ a\GS + b\GY
$$
are strongly elliptic, and
\begin{eqnarray}
\s_2(\GT)(\x)&\ge&\dfrac{n-2}{n(n+2)}|\x|^2,\label{est1} \\
\s_2(\GZ)(\x)&\ge&\dfrac{n+1}{2(n+2)}|\x|^2,\label{est2} \\
\s_2(a\; \GS + b \;\GY)(\x)&\ge&{\rm min}\left(\dfrac{a}{n}\,,
\dfrac{b(n-3)}{2(n-2)}
\right).\label{est3}
\end{eqnarray}
\end{corollary}

\begin{proof}{Proof} By the proof of Lemma \ref{strongell},
the eigenvalues $\m_i$ of $\s_2(\GT)(\x)$
for $|\x|^2=1$ must be roots of the quadratic 
$$
\begin{array}{l}
-(n+2)^2\m_i^2+\dfrac{2(n+2)(n^2-2n+2)}{n(n-2)}\m_i-1 \\
=-(n+2)^2\left(\m_i-\dfrac{n-2}{n(n+2)}\right)\left(\m_i-\dfrac{n}{(n+2)(n-2)}
\right).
\end{array}
$$
In particular, we have the estimate \nnn{est1}.
We may also compute that, in the notation of \nnn{Tws} above,
$$
\begin{array}{l}
\s_2(\GS)(\x)=\dfrac1{n(n-1)}T(\x)=
\dfrac{n|\x|^2-(n+2)(n-2)\s_2(\GT)(\x)}{4(n-1)}\,, \\
\s_2(\GY)(\x)=\dfrac{n-3}{8(n-1)}\left\{-(n-2)|\x|^2+n(n+2)\s_2(\GT)(\x)
\right\}, \\
\s_2(\GZ)(\x)=\dfrac18\left\{(n+2)|\x|^2-n(n-2)\s_2(\GT)(\x)\right\}.
\end{array}
$$
Thus, with respect to the block diagonalization in which
\beq\label{block}
\s_2(\GT)(\x)={\rm diag}\left(\dfrac{n}{(n+2)(n-2)}\,,\,\dfrac{n-2}{n(n+2)}
\right)|\x|^2,
\eeq
we also have
\beq\label{blocks}
\begin{array}{l}
\s_2(\GS)(\x)={\rm diag}\left(0,\dfrac1{n}\right)|\x|^2, \\
\s_2(\GY)(\x)={\rm diag}\left(\dfrac{n-3}{2(n-2)}\,,\,0\right)|\x|^2, \\
\s_2(\GZ)(\x)={\rm diag}\left(\dfrac{n+1}{2(n+2)}\,,\,\dfrac{n}{n+2}\right)
|\x|^2.
\end{array}
\eeq
The estimates (\ref{est2},\ref{est3}) follow.
\end{proof}

The following is a provisional form of Theorem \ref{weitz} below:

\begin{corollary}\label{provweit} The following operators have order
zero:
$$
\begin{array}{l}
Z_1:=\dfrac{(n-3)(n-2)}{2n}\;\GS-\dfrac{(n-3)(n+2)}{2n}\;\GT+\GY, \\
Z_2:=-\dfrac{(n-1)(n+2)}{2n}\;\GS-\dfrac{(n-2)(n+1)}{2n}\;\GT+\GZ.
\end{array}
$$
\end{corollary}

\begin{proof}{Proof} 
Equations (\ref{block},\ref{blocks})
show that the $Z_i$ have order at most $1$.  But invariant theory shows
that any equivariant operator of homogeneity $2$ 
and order $<2$ is an action of
the Riemann curvature.
\end{proof}

\section{Bundles associated to representations of 
the spin group}\setcounter{equation}0 
Here we would like to provide some background and motivation
for the decompositions above.
Strictly
speaking, this material is not needed to follow the arguments of this
paper.
Accordingly, we do not fill in the details of, for example,
the process of matching dominant weight labels to 
tensor symmetry types of bundles; see \cite{sw,strich} more information
along these lines.
We believe, however, that an understanding
of the representation theoretic thinking behind this work will
be valuable in further investigations.

Irreducible representations of $\spin(n)$, $n\ge 2$,
are parameterized
by {\it dominant weights} 
$(\l_1\,,\ldots\,,\l_\h)\in\bbZ^\h\cup(\frac12+\bbZ)^\h$, 
$\h=[n/2]$, satisfying the inequality constraint
\bdm
\begin{array}{ll}
\l_1\ge\ldots\ge\l_\h\ge 0,\qquad & n\text{ odd}, \\
\l_1\ge\ldots\ge\l_{\h-1}\ge|\l_\h|\,,\qquad & n\text{ even}.
\end{array}
\edm
The dominant weight $\l$ is the highest weight of the corresponding
representation.  The representations which factor through $\so(n)$ are
exactly those with $\l\in\bbZ^\h$.  We shall denote by $V(\l)$ the
representation with highest weight $\l$.  If $M$ is an $n$-dimensional
smooth manifold with $\spn$ structure and $\cF$ is the bundle of spin
frames, we denote by $\bbV(\l)$ the vector
bundle $\cF\times_\l V(\l)$.  When spin structure is not involved
(i.e.\ when $\l$ is integral), we may use the orthonormal frame bundle
in constructing $\bbV(\l)$. 

One important highest weight is that of
the defining representation $V(1,0,\ldots,0)$ of $\so(n)$.  The
classical {\it selection rule} describes the $\spin(n)$ decomposition
of\newline 
$V(1,0,\ldots,0)\otimes V(\l)$ for an arbitrary dominant $\l$:
\bdm
V(1,0,\ldots,0)\ot V(\l)\cong_{\spin(n)}V(\s_1)\os\ldots\os V(\s_{N(\l)}),
\edm
where the $\s_u$ are {\em distinct}: $\s_u\cong_{\spin(n)}\s_v\Rightarrow
u=v$. A given $\s$ appears 
if and only if
$\s$ is a dominant weight and 
\begin{equation}\label{selection}
\begin{array}{l}
\s=\l\pm e_a\,,\text{ some }a\in\{1,\ldots,\h\},\qquad 
\underline{\text{or}} \\
n\text{ is odd, }\l_\h\ne 0,\ \s=\l.
\end{array}
\end{equation}
Here $e_a$ is the $\ath$ standard basis vector in $\bbR^\h$.
Note that 
$N(\l)$, the number of selection rule ``targets'' of $V(\l)$, depends on $\l$.
We shall use the notation
$$
\l\lra\s
$$
for the selection rule: $\l\lra\s$ if and only if $V(\s)$ appears 
as a summand in\newline 
$V(1,0,\ldots,0)\ot V(\l)$.
The notation $\lra$ is justified because the relation is symmetric.
In fact, one can see {\it a priori} that the relation must be symmetric:
the defining representation of $\so(n)$ is real, 
and thus self-contragredient.

An interesting concept related to the selection rule is that of {\em
generalized gradients}, or {\it Stein-Weiss operators} \cite{sw}.
The covariant derivative $\N$ carries sections of
$\bbV(\l)$ to sections of 
\bdm
\begin{array}{rl}
T^*M\ot\bbV(\l)&\cong_{\spin(n)}\bbV(1,0\ldots,0)\ot\bbV(\l) \\
&\cong_{\spin(n)}\bbV(\s_1)\os\ldots\os\bbV(\s_{N(\l)}).
\end{array}
\edm
Since the selection rule is 
multiplicity free, we may project
onto the unique $\s_u$ summand; the result is our gradient:
$$
G_u=G_{\l\s_u}={\bf P}_u\circe\N.
$$
Up to normalization and isomorphic realization of bundles, 
some examples of gradients, or direct sums of
gradients, are the exterior
derivative $d$, its formal adjoint $\d$, the conformal Killing
operator $S$, the Dirac operator, the twistor operator, and the
Rarita-Schwinger operator.  In fact, every first-order
$\spin(n)$-equivariant differential operator is a direct sum of
gradients \cite{feg}.

In even dimensions, if $\l_\h\ne 0$, the bundles $\bbV(\l)$ and
$\bbV(\bar\l)$, where
$$
\bar\l=(\l_1\,,\ldots,\l_{\h-1},-\l_\h),
$$
are distinguished by the action of 
the volume element, which commutes with the action of SO$(n)$ or
Spin$(n)$, but not with that of O$(n)$ or Pin$(n)$.
This shows up in
{\em duality} and {\em chirality} considerations.  For example,
we get the split between the two eigenbundles of the pointwise operator 
$E_{\a_1\,\ldots\a_n}\g^{\a_1}\cdots\g^{\a_n}$ on spinors,
or between the two eigenbundles of the Hodge $\star$ operator on
(complexified) $n/2$-forms.  In the situation where duality and/or 
chirality are not in play, it is convenient to define,
for $\l_\h\ge 0$,
\bdm
\bbU(\l):=\left\{\begin{array}{ll}
\bbV(\l)\os\bbV(\bar\l),\ \ &n\ {\rm even\ and}\ \l_\h>0, \\
\bbV(\l) &{\rm otherwise.}
\end{array}\right.
\edm
The bundle $\bbU(\l)$ is defined when 
$$
\l_1\ge\ldots\ge\l_\h\ge 0
$$
for both even and odd $n$.
When $\l_\h=1/2$ (a case of much interest in the present work), 
$\bbU(\l)$ is always a direct summand in $T^*M\ot\bbU(\l)$.

The gradient concept is
the motivation behind the definitions of the operators
$\gS\,$, $\gT\,$, $\gY\,$, and $\gZ$ above.  The spinor bundle is
$$
\Sp\cong_{\spn}\bbU(\tf\,,\ldots,\tf).
$$
Similarly, the twistor bundle is
\bdm
\Tw\cong_{\spn}\bbU(\tfrac32\,,\tf\,,\ldots,\tf).
\edm
The other bundles introduced in the last section are
\bdm
\Yb\cong_{\spn}
\left\{\begin{array}{ll}
\bbU(\tfrac32\,,\tfrac32\,,\tf\,,\ldots,\tf),\ \ &n\ge 4, \\
\bbU(\tf), &n=3, \\
0, &n=2,
\end{array}\right.
\edm
and
$$
\Zb\cong_{\spn}
\bbU(\tfrac52\,,\tf\,,\ldots,\tf).
$$

The selection rule \nnn{selection} shows that there are gradients, or
direct sums of gradients, acting between copies of the following bundles.
Up to isomorphic realizations of bundles and constant factors,
the corresponding gradients or direct sums of gradients are also listed:
$$
\begin{array}{ll}
\Sp\to\Sp\qquad & \dc \\
\Sp\to\Tw\qquad & \bT \\ 
\Tw\to\Sp\qquad & \bT^*\ {\rm or}\ \gS \\
\Tw\to\Tw\ (n\ge 3)\qquad & \cS^0\ {\rm or}\ \gT \\ 
\Tw\to\Yb\ (n\ge 4)\qquad & \gY \\ 
\Tw\to\Zb\qquad & \gZ
\end{array}
$$
(Recall the definition of $\cS^0$ in \nnn{rar1}.) 
The operators $\bT^*$ and $\gS$ are targeted at different realizations of
$\Sp$; thus they are not the same operator, but each is 
a constant factor times
the composition of the other with a $\spn$-bundle isomorphism.
A similar statement holds for 
$\cS^0$ and $\gT$ with $\Tw\,$. More precisely, using
\nnn{isoinj},
\nnn{egS}, and 
\nnn{egT}, one has 
\beq\label{ident}
\begin{array}{ll}
\bT^* & = \frac12\; I_{{\bf\S}}^*\;\gS  \\
\cS^0 & = - I_{{\bf T}}^*\;\gT\,.
\end{array}
\eeq

Certain natural first-order operators are especially interesting
in that they can be realized so that the source and target bundle
are the same.  In particular, these operators have spectra.
Examples are the Dirac and Rarita-Schwinger operators, and the
operator $\star d$ on $\G(\L^{(n-1)/2})$ for odd $n$.  Such operators
arise as follows.  In odd dimensions, we take the gradient corresponding
to the exceptional case of the selection rule (the second line of
\nnn{selection}).
In this case, $G_{\l\l}$ carries sections of $\bbV(\l)$ to sections of
a copy
of $\bbV(\l)$ which lives as a subbundle in $T^*M\otimes\bbV(\l)$.
If we would like to use the source
realization of $\bbV(\l)$ as both source and target for
a realization $\dls$ of $G_{\l\l}$, 
we need a choice of normalization.
First, normalize the Hermitian inner product on $T^*M\otimes\bbV$
so that 
\begin{equation}\label{nmconv}
|\xi\otimes v|^2=|\xi|^2|v|^2;
\end{equation}
then normalize $\dls$ so that 
\beq\label{dsnormodd}
\dls^2=\gll^*\gll.
\eeq
This determines
$\dls$ up to multiplication by $\pm 1$.  

In even dimensions, take a dominant weight $\l$ with $\l_\h=1/2$.
Then there are gradients
$G_{\l\bar\l}$ and $G_{\bar\l\l}\,$, giving rise to a 
first-order operator 
\bdm
\left(\begin{array}{cc}
0 & G_{\bar\l\l} \\
G_{\l\bar\l} & 0
\end{array}\right)
\edm
carrying sections of $\bbU(\l)=\bbV(\l)\os\bbV(\bar\l)$ to sections of 
an isomorphic copy of this bundle,
realized in its tensor product with $T^*M$.
Remarks similar to those above, on
normalization and realization, then hold, and we obtain a 
first-order differential operator 
$\cS_{\l\os\bar\l}$ on $\G(\bbV(\l)\os\bbV(\bar\l))$,
again determined up to a factor of $\pm 1$, normalized so that 
\beq\label{dsnormeven}
(\cS_{\l\os\bar\l})^2=G_{\l\bar\l}^*G_{\l\bar\l}
\os G_{\bar\l\l}^*G_{\bar\l\l}\,.
\eeq

The sign ambiguity in the $\cS$ operators 
is in the nature of things: it is analogous to the
ambiguity in the naming of the complex units $\pm\eye$.  Indeed, this
is more than an analogy: gradients generalize the Cauchy-Riemann
equations \cite{sw}, which are sensitive to the renaming of $\pm\eye$.
In our examples, the ambiguity may be viewed as residing in a choice
of fundamental tensor-spinor (or more generally, a Clifford structure)
$\g$.  The Clifford relations and spin connection (in particular the
relation $\N\g=0$) are invariant under interchange of $\g$ and $-\g$,
but the Dirac operator $\g^a\N_a$ undergoes a sign change.  

The principal examples of such {\em self-gradients} of interest
to us are those which act on $\G(\Sp)$ and $\G(\Tw)$, namely
\bdm
\cS_{{\bf\S}} = \frac{1}{\sqrt{n}}\;\dc
\edm
and
\bdm
\cS_{{\bf T}} : \f\mapsto\sqrt{\dfrac{n}{(n+2)(n-2)}}\;\left(
\g^\l\N_\l\f_\a-\dfrac2n\g_\a\N^\l\f_\l\right)\,,
\edm
i.e.,
\bdm
\begin{array}{l}
\cS_{{\bf T}}= \sqrt{\dfrac{n}{(n+2)(n-2)}}\;\cS^0
= - \sqrt{\dfrac{n}{(n+2)(n-2)}}\;I^*_{{\bf T}}\; \gT\,.
\end{array}
\edm

The normalizations are computed from \nnn{dsnormodd} and \nnn{dsnormeven}.
The {\em normalized Rarita-Schwinger operator} $\cS_{{\bf T}}$ will appear in
formulas below.
We shall also have use for self-gradients on $\Yb$ and $\Zb\,$; see
\nnn{sgsf2} and \nnn{selfZ} below.

An important point is that there are distinguished normalizations
for $\GS$, $\GT$, $\GY$, $\GZ$, and all similarly defined operators.
In fact, the issue is exactly that of normalizing the formal adjoint
by getting a relative normalization for the source and target bundles
of a gradient (or a suitable direct sum of gradients), 
say $\bbV$ and $\bbW$ respectively.  This is provided by taking the 
realization of $\bbW$ in $T^*M\ot\bbV$, and normalizing its metric
according to \nnn{nmconv}.  
Allowing the metric on $\bbV$ to determine that on $\bbW$ in this way,
our operators $G^*G$
remain invariant under 
rescalings of the metric on $\bbV$.

\section{Bochner-Weitzenb\"ock formulas}\setcounter{equation}0
A {\em Bochner-Weitzenb\"ock formula} (henceforth a
{\em BW formula}) may be described in general as an equation
$$
D_1=D_2+Z,
$$
where $D_1$ and $D_2$ are natural, nonnegative definite second-order
differential operators with the same leading symbol, on sections of a
vector bundle ${\bf V}$, and $Z$ is a natural bundle 
endomorphism of ${\bf V}$; in
particular, a differential operator of order zero.  The importance of
such formulas derives
from the elementary observation that if $Z\ge
c\cdot\id_{\bf V}$ pointwise, for some constant $c>0$, then $D_1\ge
c\cdot\id_{\G({\bf V})}\,$; if $-Z\ge k\cdot\id_{\bf V}\,$,
then $D_2\ge k\cdot\id_{\G({\bf V})}\,$.  
This observation usually appears as part of a
longer argument in which devices particular to the situation, notably
variations of the underlying geometric structure, are also employed.

There is an essentially unique BW formula on $\Sp\,$.  One way to 
express this is the {\em Schr\"odinger-Lichnerowicz formula} \cite{lich}
\beq\label{lichfm}
\dc^2=\N^*\N+\dfrac K4,
\eeq
where $K$ is the scalar curvature.
By the discussion at the end of the last section, 
$$
\N^*\N= \cS_{{\bf\S}}^2+\bT^*\bT = \dfrac1n\dc^2+\bT^*\bT,
$$
so \nnn{lichfm} is equivalent to
\beq\label{lichid}
\dc^2=\dfrac n{n-1}\bT^*\bT+\dfrac{nK}{4(n-1)}\,;
\eeq
this is sometimes known as the {\em Lichnerowicz identity} \cite{lich2}.  It
is, in a certain sense, an optimal way to write \nnn{lichfm}, since it brings
us into contact with an {\em orthogonal} decomposition of $\N\y$.

Both \nnn{lichfm} and \nnn{lichid} are manifestations of the same BW
formula: each computes the same linear combination of 
$\dc^2$ and $\bT^*\bT$; the unique such combination 
that has vanishing leading symbol.
As Corollary \ref{provweit} makes clear, 
the operators $\GS$, $\GT$, $\GY$, and $\GZ$ give rise to two
essentially different BW formulas.
To state these, let us standardize some notation.  Let 
$R$ be the Riemann curvature tensor, with the convention on
index placement that gives $[\N_\a\,,\N_\b]X^\l=
R^\l{}_{\m\a\b}X^\m$ for $X$ a vector field.  Then $r_{\m\b}=
R^\a{}_{\m\a\b}$ is the Ricci tensor, and $K=r^\b{}_\b$ is the 
scalar curvature.  Define the {\em Einstein} (trace-free Ricci) {\em tensor}
$b$ by
\beq\label{ein}
b_{\a\b}=r_{\a\b}-\dfrac Kng_{\a\b}\,,
\eeq
and define the {\em Weyl tensor} $C$ to be 
the totally trace-free part of $R$.  Explicitly, if
$$
J:=\dfrac{K}{2(n-1)}\qquad {\rm and}\qquad 
V:=\dfrac{r-Jg}{n-2}\,,
$$
then
\beq\label{wl}
C^\a{}_{\b\kappa\l}=R^\a{}_{\b\kappa\l}+V_{\b\kappa}\d^\a{}_\l
-V_{\b\l}\d^\a{}_\kappa
+V^\a{}_\l g_{\b\kappa}-V^\a{}_\kappa g_{\b\l}\,.
\eeq

If $\f\in\G(\Tw)$, let
\bdm
\begin{array}{rl}
(b\cdot\f)_\m:&=\dfrac1n\left((n-2)b_{\m\l}\f^\l-b_{\a\l}\g^\a\g_\m\f^\l
\right) \\
&=\dfrac1n\left((n-1)b_{\m\l}\f^\l-b_{\a\l}\g^\a{}_\m\f^\l
\right), \\
(C\diamond\f)_\m:&=C_{\a\b}{}^\l{}_\m\g^\a\g^\b\f_\l=
C_{\a\b}{}^\l{}_\m\g^{\a\b}\f_\l\,.
\end{array}
\edm
By the Clifford relations, the Bianchi identity, 
and the trace-free nature of $C$,
\bdm
\g^\m(b\cdot\f)_\m=\g^\m(C\diamond\f)_\m=0.
\edm
Thus both $b\,\cdot$ and $C\diamond$ are $\spn$-bundle 
endomorphisms of $\Tw\,$.
In fact, the maps $\b\ot\f\mapsto\b\cdot\f$ and $\z\ot\f\mapsto\z\diamond\f$
describe $\spn$-bundle homomorphisms $\tfs^2\ot\Tw\to\Tw$ and
$\cW\ot\Tw\to\Tw$ respectively, where 
$\cW$ is the bundle
of algebraic Weyl tensors.  Note that by \cite{strich},
$$
\tfs^2\cong_{\spn}\bbV(2,0,\ldots,0),
$$
and
\bdm
\cW\cong_{\spn}\bbU(2,2,0,\ldots,0)=\left\{\begin{array}{ll}
\bbV(2,2,0,\ldots,0), &n\ge 5, \\
\bbV(2,2)\os\bbV(2,-2),\ \ &n=4, \\
0, &n<4.
\end{array}\right.
\edm

By the Clifford relations, the skew-adjointness of $\g^\m$,
and the twistor condition $\g^\m\f_\m=0$, we have
\beq\label{bac}
\langle(b\cdot\f)_\m\,,\f^\m\rangle=b_\m{}^\l\langle\f_\l\,,\f^\m
\rangle.
\eeq
\begin{theorem}\label{weitz}
\bdm
\begin{array}{rl}
Z_1:&=\dfrac{(n-3)(n-2)}{2n}\;\GS-\dfrac{(n+2)(n-3)}{2n}\;\GT+\GY \\
&=\dfrac18\,C\diamond
+\dfrac{n-3}{2(n-2)}\;b\cdot-\dfrac{(n-2)(n-3)}{8n(n-1)}\,K,
\end{array}
\edm
\bdm
\begin{array}{rl}
Z_2:&=-\dfrac{(n-1)(n+2)}{2n}\;\GS-\dfrac{(n+1)(n-2)}{2n}\;\GT+\GZ \\
&=\dfrac38\,C\diamond
-\dfrac{n+1}{2(n-2)}\;b\cdot-\dfrac{(n+2)(n+1)}{8n(n-1)}\,K.
\end{array}
\edm
\end{theorem}

\begin{proof}{Proof} We promote the leading symbol calculation
of Lemma \ref{strongell} and Corollaries \ref{eigrk} and \ref{provweit}
to operator calculations, keeping track of curvature terms.
The following identities are used.  Let $\cR$ be the Riemannian spin
curvature; that is, $\cR_{\a\b}=[\N_\a\,,\N_\b]$, the precise
effect of which depends on what sort of index expression appears to its right.
If $\psi$ is a spinor, then
$$
\cR_{\l\m} \psi := W_{\l\m} \psi = -\frac14R_{\kappa\nu\l\m}\g^\kappa\g^\nu
\psi,
$$
where $W$ is the spin curvature.
If $\f$ is a spinor-one-form, then
$$
\cR_{\l\m}\f_\a=W_{\l\m}\f_\a-R^\nu{}_{\a\l\m}\f_\nu\,.
$$
The classical Lichnerowicz calculation, which combines the Clifford relations
and the Bianchi identity, shows that
$$
\g^\m W_{\l\m}=-\frac12r_{\l\m}\g^\m.
$$
In particular,
$$
\g^\l\g^\m W_{\l\m}=\frac12K.
$$
This leads to the formula
$$
\g^\l\g^\m\cR_{\l\m}\f_\a=4W^\l{}_\a\f_\l+\frac12K\f_\a\,.
$$
We also have
$$
\cR^\a{}_\m\f_\a=(W^\a{}_\m+r^\a{}_\m)\f_\a\,,
$$
so that
$$
\g^\m\cR^\a{}_\m\f_\a=\frac12r^\a{}_\m\g^\m\f_\a\,.
$$
We then write the curvature terms in terms of $K$, $b$, and $C$ using
(\ref{ein},\ref{wl}).
\end{proof}

\begin{remark}\label{genbw} {\rm 
In \cite{tbsw}, Theorem 5.10, it is shown that, for the $N(\l)$
gradients $G_u$ 
emanating from a given irreducible $\spn$-bundle $\bbV(\l)$,
$$
\dim\spam\Big\{\s_2(G_u^*G_u)\mid u=1,\cdots,N(\l)\Big\}=\left[\dfrac{N(\l)+1}2
\right]\,.
$$
That is, only about half of the leading symbols of the $G_u^*G_u$
are linearly independent.  This means that there is an
$[N(\l)/2]$-parameter family of BW formulas relating these operators
(since $[N(\l)/2]=N(\l)-[(N(\l)+1)/2]$).  In fact, there are numbers
$\tilde c_u$ and $s_u$ such that 
\beq\label{cusu}
\s_2\left(\sum_{u=1}^{N(\l)}b_uG_u^*G_u\right)=0
\iff\sum_{u=1}^{N(\l)}b_u\tilde c_us_u^{2j}=0,\ \ j=0,1,\cdots,\left[
\dfrac{N(\l)+1}2\right]-1.
\eeq
Given such a ``null'' linear combination of operators $G_u^*G_u\,$,
the corresponding BW formula takes the form
$$
\sum_{b_u>0}b_uG_u^*G_u=\sum_{-b_u>0}(-b_u)G_u^*G_u+Z,
$$
where $Z$ has order zero.}
\end{remark}

\begin{remark}\label{twospecialbw} {\rm 
Among all the BW formulas described in the last remark are some
distinguished ones.  First, for any $\bbV(\l)$, there is a formula,
studied by Gauduchon in \cite{gau}, Appendice B, in which the order zero
operator $Z$ is controlled by the curvature operator, in an appropriate
sense.  The value of the {\em Casimir operator} of the Lie algebra 
$\gso(n)$ in 
the representation $V(\l)$ is, in the notation of the last section,
\beq\label{casval}
\l(\cason)=\sum_{a=1}^\h\left\{\left(\l+\dfrac{n-2a}2\right)^2-
\left(\dfrac{n-2a}2\right)^2\right\}.
\eeq
Let
\beq\label{defsu}
s_u:=\dfrac12\left(\l(\cason)-\s_u(\cason)\right);
\eeq
this is in fact the same quantity $s_u$ that appears in \nnn{cusu}.
Gauduchon showed that if $n\ge 3$ and $\l$ is integral,
\beq\label{gauresult}
\cP_\l:=\sum_{u=1}^{N(\l)}\left(s_u+\dfrac{n-1}2\right)G_u^*G_u
=-\sum_I\l(X_I)\l(\Rop(X_I)),
\eeq
where $(X_I)$ is any local orthonormal frame for the bundle $\L^2M$,
and $\Rop$ is the curvature operator on sections $\eta$ of $\L^2M$:
$$
(\Rop\eta)_{\a\b}=\df R_{\a\b}{}^{\m\nu}\eta_{\m\nu}\,.
$$
(The fact that this particular linear combination of $G_u^*G_u$
must produce a BW formula was actually noted earlier in \cite{cph};
see
Remark \ref{kbh} below.) 
In \cite{improv}, the present authors show, among other things, 
that this result extends to half-integral $\l$.  
For any $\l$, the significance of the result is that it enforces
pointwise bounds on
$\cP_\l\,$, which has order zero as a differential operator and thus
is actually a section of $\endo\,\bbV(\l)$.  These bounds 
are multiples of the bottom and top pointwise eigenvalues
of the curvature 
operator $\Rop$, which is a section of $\endo\,\L^2$.  
If, at a point
$x\in M$,
\beq\label{RopBounds}
q_x\;\id_{\L^2_xM}\le\Rop_x\le Q_x\;\id_{\L^2_xM}
\eeq
for some constants $q_x\,,\,Q_x\,$,
in the sense of ordering of endomorphisms ($A\le B$ iff $B-A$ is positive
semidefinite), then
\beq\label{CasBounds}
q_x\;\l(\cason)\;\id_{\bbV(\l)_x}\le (\cP_\l)_x \le
Q_x\;\l(\cason)\;\id_{\bbV(\l)_x}\,,
\eeq
also in the sense of endomorphism ordering.  (Note that both
endomorphisms, $\Rop$ and $(\cP_\l)_x\,$, are symmetric.)

Another important combination of the $G_u^*G_u$ is the one 
involved in the conformally
covariant operator \cite{cph,pspum}
$$
\cD_\l=\dfrac{K}{2(n-1)}-\sum_{u=1}^{N(\l)}\left(s_u+\dfrac12\right)^{-1}
G_u^*G_u\,.
$$
This is well-defined as long as no $s_u+\tf$ vanishes; i.e., provided
it is {\em not} the case that
\beq\label{excDl}
n\ \mbox{is even and}\ \l_\h=0\ne\l_{\h-1}\,.
\eeq
When \nnn{excDl} does not hold,
\bdm
\ord\,\cD_\l=\left\{\begin{array}{ll}
2,\qquad & N(\l)\ {\rm odd,} \\
0,\qquad & N(\l)\ {\rm even.}
\end{array}\right.
\edm
In case $\ord\,\cD_\l=0$, we have a BW formula; by the conformal
covariance of $\cD_\l$ and some invariant theory,
$$
\cD_\l\f=\a(C,\f)
$$
for all sections $\f$ of $\bbV(\l)$, where $\a$ is some section
of the bundle $\hom(\cW\ot\bbV(\l),\bbV(\l))$.  In other words,
the BW formula that we get from the combination of $G_u^*G_u$
appearing in $\cD_\l$ {\em omits} the Einstein tensor, in the sense
that its right-hand side depends only on $K$ and $C$.
This formula was exploited in \cite{bh} to set up a systematic approach for
obtaining vanishing theorems, and in some cases
eigenvalue estimates, based on the relative size of the bottom
eigenvalue of the Yamabe operator $\D+(n-2)K/(4(n-1))$ on scalars,
and the pointwise eigenvalues of the Weyl tensor $C$.
To be precise, the relevant Weyl tensor 
data are bounds of the form
\nnn{RopBounds} on
the operator $\Cop$ defined in analogy with the curvature operator:
$(\Cop\eta)_{\a\b}=\frac12C_{\a\b}{}^{\m\nu}\eta_{\m\nu}\,$.

When $N(\l)\ge 4$ is even and \nnn{excDl} does not hold, the BW formulas
associated to the $\cP_\l$ and $\cD_\l$ are different; that is, the 
coefficient arrays
\beq\label{coefarrays}
\left(s_u+\dfrac{n-1}2\right)_{u=1}^{N(\l)}\qquad{\rm and}\qquad\left(
\left(s_u+\df\right)^{-1}\right)_{u=1}^{N(\l)}
\eeq
are linearly independent (\cite{improv}, Lemma 2.1).  
When $N(\l)=2$, they must be proportional by the above-described result
counting the linearly independent BW formulas.  In fact, when $N(\l)=2$,
the first array in \nnn{coefarrays} is 
$2\l(\cason)/n$ times the
second (\cite{improv}, Remark 2.2).
}
\end{remark}

\begin{remark}\label{kbh} {\rm 
In \cite{cph}, p.46, equation (3.30), an interesting relation between
$\cP_\l$ and $\cD_\l$ is noted.  This holds in all cases except
\nnn{excDl}; i.e., even if $\cD_\l$ has order 2.  Up to a nonzero
constant multiple, the second conformal variation of $\cD_\l$ in the
direction of conformal factors $e^{2\e\w}$, where $\w\in\ci(M)$, and
$\e\in\bbR$ is a variational parameter, is the second-order symbol
of $\cP_\l$ evaluated at the covector field $\x=d\w$.  Since $\cD_\l$
is conformally covariant and $(d\w)_x$ may be arbitrarily prescribed
at any point $x\in M$, this establishes that $\cP_\l$ has order less
than 2, and thus (by invariant theory and homogeneity) 
order 0.  This argument, however,
does not provide the precise right-hand side of \nnn{gauresult}.
}
\end{remark}

For $n\ge 4$, the twistor bundle $\Tw$ is either a $\spn$-irreducible
bundle with $N(\l)=4$ ($n$ odd), or a direct sum of two 
$\spn$-irreducibles, each having $N(\l)=4$.  Thus the list of BW
formulas in Theorem \ref{weitz} above is complete.  Since there
are exactly two BW formulas, the last remark shows that each is
a linear combination of the formulas associated to the 
$\cP_\l$ and the $\cD_\l$ combinations.  In fact, the conformally
covariant operator (or direct sum of such for $n$ even) is
\beq\label{TBop}
\begin{array}{rl}
\cD&=\dfrac{K}{2(n-1)}-\dfrac2{n+1}\GS-2\GT\\
&\qquad+\dfrac2{n-3}\GY + \dfrac2{n+1}\GZ \\
&=\dfrac{K}{2(n-1)}+\dfrac2{n-3}Z_1+\dfrac2{n+1}
Z_2 \\
&=\dfrac{n-2}{(n-3)(n+1)}\,C\diamond\,.
\end{array}
\eeq
The Gauduchon operator (or direct sum of such if $n$ is even) is
\beq\label{GauTwi}
\begin{array}{rl}
\cP&=\dfrac{2n-1}2\GS+\dfrac{n-1}2\GT \\
&\qquad+\df\GY-\dfrac32\GZ \\
&=\df Z_1-\dfrac32Z_2 \\
&=-\df\,C\diamond+\dfrac{n}{n-2}\,b\cdot+\dfrac{n+7}{8(n-1)}\,K \\
&=:R\diamondsuit\,.
\end{array}
\eeq
The $R\diamondsuit$ notation will be useful below.
One can show control over the curvature action $R\diamondsuit$ by the 
curvature operator by an elementary
argument, without using representation theory.  Let $A$ be the following
action of two-forms on twistors:
$$
(A(\eta)\f)_\m=\eta_{\a\b}\g^\a\g^\b\f_\m-4\eta_\m{}^\l\f_\l\,.
$$
A short calculation shows that $\g^\m(A(\eta)\f)_\m=0$, so 
$A\in\hom(\L^2M\ot\Tw,\Tw)$.  Another short calculation shows that
$A(\eta)$ is skew-adjoint as a bundle endomorphism of $\Tw\,$.
Thus if $(X_I)$ is an orthonormal basis of $\L^2_xM$ diagonalizing
the symmetric operator $\Rop_x$ with eigenvalues $a_I\,$, we have
\beq\label{Ac*Ac}
-\sum_Ia_IA(X_I)^2
\eeq
nonnegative (or nonpositive) if $\Rop$ is nonnegative (or nonpositive).
However, direct calculation shows that
$$
-\sum_Ia_IA(X_I)^2=8R\diamondsuit\,.
$$
(One can organize the calculation as follows: compute $-A(\eta)^2\f$,
and then replace each occurrence of $\eta_{\a\b}\eta_{\l\m}$ by
$\frac12R_{\a\b\l\m}\,$; this is the value of \nnn{Ac*Ac}.)
In particular,
$$
\begin{array}{l}
\Rop\ge 0\qquad\implies\qquad R\diamondsuit\ge 0, \\
\Rop\le 0\qquad\implies\qquad R\diamondsuit\le 0.
\end{array}
$$

To get precise control by the curvature operator, let $[[K]]$, $[b]$, and
$C$ be the (orthogonal) contributions of the scalar curvature $K$,
the Einstein tensor $b$, and the Weyl tensor $C$ to the Riemann curvature $R$:
$$
R=C+[b]+[[K]].
$$
In detail,
$$
\begin{array}{rl}
[[K]]_{\a\b\l\m}&=\dfrac{K}{n(n-1)} \left(g_{\a\l}g_{\b\m}-g_{\a\m}g_{\b\l}
\right), \\
{[}b]_{\a\b\l\m}&=
\dfrac1{n-2} \left(b_{\b\m}g_{\a\l}-b_{\b\l}g_{\a\m}-b_{\a\m}g_{\b\l}
+b_{\a\l}g_{\b\m}\right).
\end{array}
$$
The curvature operator is
$$
(\Rop\eta)_{\a\b}=\dfrac1{n(n-1)}\,K\eta_{\a\b}
+\dfrac1{n-2}\left(b^\l{}_\a\eta_{\l\b}+b^\l{}_\b\eta_{\a\l}\right)
+\df C_{\a\b}{}^{\l\m}\eta_{\l\m}\,.
$$
Let $q$ and $Q$ be smooth real-valued functions on $M$.  
The condition that $\Rop\ge q\id_{\L^2M}$ is thus equivalent to the condition
that the algebraic curvature tensor
\beq\label{rq}
R_q:=[[K-n(n-1)q]]+[b]+C
\eeq
has its curvature operator $\Rop_q$ 
nonnegative.  Similarly, the condition $\Rop\le Q\,\id_{\L^2M}$
is equivalent to the condition that $\Rop_Q$ is nonpositive.
We have:
\beq\label{RopandRq}
\begin{array}{c}
q\le\Rop\le Q\qquad \iff\qquad\Rop_Q\le 0\le\Rop_q\qquad\implies \\
R_Q\diamondsuit\le 0\le
R_q\diamondsuit\qquad \iff\qquad\dfrac{n(n+7)}{8}\,q\le R\diamondsuit\le
\dfrac{n(n+7)}{8}\,Q \\
\qquad \iff\qquad\dfrac{n(n+7)}{8}\,q\le\cP\le
\dfrac{n(n+7)}{8}\,Q.
\end{array}
\eeq
Here we have used the fact that, by \nnn{GauTwi} and \nnn{rq},
$$
R_f\diamondsuit=R\diamondsuit-\dfrac{n(n+7)}8f.
$$
To compare with \nnn{CasBounds}, note that the value of the Casimir
operator in the twistor representation(s) is $n(n+7)/8$.

The same reasoning may be applied to 
get precise control over the quantity in \nnn{TBop}.  First notice
that 
$$
C\diamond=-2C\diamondsuit.
$$
If for some smooth functions $t,T$ we have
$$
t\le\Cop\le T,
$$
then
$$
\dfrac{n(n+7)}{8}\,t\le C\diamondsuit\le
\dfrac{n(n+7)}{8}\,T,
$$
whence
$$
-\dfrac{n(n+7)}4\,T\le C\diamond\le-\dfrac{n(n+7)}4\,t,
$$
and
$$
\dfrac{(n-2)n(n+7)}{4(n-3)(n+1)}\,t+\dfrac{K}{2(n-1)}
\le \;\cD_0\;\le
\dfrac{(n-2)n(n+7)}{4(n-3)(n+1)}\,T+\dfrac{K}{2(n-1)}\,,
$$
where
\beq\label{elimb}
\begin{array}{rl}
\cD_0:&=
\dfrac2{n+1}\GS+2\GT \\
&\qquad-\dfrac2{n-3}\GY-\dfrac2{n+1}\GZ \\
&=\dfrac{K}{2(n-1)}-\dfrac{n-2}{(n-3)(n+1)}C\diamond\,.
\end{array}
\eeq

Thus if $q,Q,t,T$ are as above,
$$
\begin{array}{c}
\dfrac{2n-1}2\;\GS+\dfrac{n-1}2\;\GT+\df\GY\ge\cP\ge\dfrac{n(n+7)}8q, \\
\dfrac32\GZ\ge-\cP\ge-\dfrac{n(n+7)}8Q, \\
\dfrac2{n+1}\;\GS + 2 \GT\ge\cD_0\ge\dfrac{(n-2)n(n+7)}{4(n-3)(n+1)}\;t
+\dfrac{K}{2(n-1)}\,, \\
\dfrac2{n-3}\;\GY+\dfrac2{n+1}\;\GZ\ge-\cD_0\ge
-\dfrac{(n-2)n(n+7)}{4(n-3)(n+1)}\;T-\dfrac{K}{2(n-1)}\,,
\end{array}
$$
as operator inequalities.  We immediately get the following
vanishing results:

\begin{theorem}\label{van} Let $n\ge 4$.
If $\Rop>0$, then $\cN(\gS)\cap\cN(\gT)\cap\cN(\gY)=0$.
If $\Rop<0$, then $\cN(\gZ)=0$.
If
$$
\Cop>-\dfrac{2(n-3)(n+1)K}{(n-1)(n-2)n(n+7)}
$$
(in particular, if $g$ is conformally flat and $K>0$), then
$\cN(\gS)\cap\cN(\gT)=0$.  If
$$
\Cop<-\dfrac{2(n-3)(n+1)K}{(n-1)(n-2)n(n+7)}
$$
(in particular, if $g$ is conformally flat and $K<0$), then
$\cN(\gY)\cap\cN(\gZ)=0$.
\end{theorem}

The BW formula \nnn{elimb} is potentially interesting because it 
eliminates the Einstein tensor term.  We may similarly compute
BW formulas that eliminate other elements: the Weyl tensor, the
scalar curvature, or any given Stein-Weiss operator ${\bf G}^*{\bf G}$.
Taking $3\cdot Z_1-Z_2\,$, we have
\beq\label{elimc}
\begin{array}{l}
\dfrac{2\left\{\left(n-\frac74\right)^2+\frac{15}{16}\right\}}{n}\;\GS
-\dfrac{\left(n-\frac12\right)^2-\frac{33}{4}}{n}\;\GT \\ 
\qquad+3 \GY-\GZ 
=2b \cdot-\dfrac{n-8}{4n}K.
\end{array}
\eeq
This eliminates the Weyl tensor $C$.  In particular, for $n=8$, one has
\beq\label{elimck}
10\GS-6\GT+3\GY-\GZ=2b \,\cdot\,;
\eeq
this eliminates both $C$ and $K$.
We eliminate $K$ by looking at 
$$
(n+1)(n+2)Z_1-(n-3)(n-2)Z_2\,.
$$
The result is
\bdm
\begin{array}{l}
(n-3)(n-2)(n+2)\GS-4(n-3)(n+1)\GT \\
\qquad\qquad+(n+1)(n+2)\GY-(n-3)(n-2)\GZ \\
\qquad=-\dfrac{(n-8)(n-1)}{4}C\diamond+\dfrac{(n-3)n(n+1)}{n-2}b\cdot\,.
\end{array}
\edm

We eliminate $\GS$ by looking at
$$
(n-1)(n+2)Z_1+(n-3)(n-2)Z_2\,.
$$
This gives
\beq\label{elimGS}
\begin{array}{l}
-(n-3)n^2\GT+(n-1)(n+2)\GY+(n-3)(n-2)\GZ \\
\qquad=\tfrac12\left\{\left(n-\frac74\right)^2+\frac{15}{16}\right\}
C\diamond+\dfrac{(n-3)n}{n-2}b\cdot-\dfrac{(n-3)(n-2)(n+2)}{4(n-1)}K.
\end{array}
\eeq

We eliminate $\GT$ by looking at
$$
(n+1)(n-2)Z_1-(n+2)(n-3)Z_2\,.
$$
This gives
\bdm
\begin{array}{l}
n^2(n-3)\GS+(n+1)(n-2)\GY-(n+2)(n-3)\GZ \\
\qquad=-\frac14\left\{(n-\tfrac12)^2-\tfrac{33}4\right\}C\diamond
+\dfrac{(n-3)n(n+1)}{n-2}b\cdot+\dfrac{(n+1)(n-3)}{n-1}K.
\end{array}
\edm

\begin{proposition}\label{bbounds}
Suppose the Einstein tensor satisfies $p\le b\le P$ in the sense of
endomorphisms, for some constants $p,P$.
(Here we view $b$ as residing in $\endo\,TM$).
If 
$$
p\ge\dfrac{n-8}{8n}K,
$$
with strict inequality at some point, 
then $\cN(\gS)\cap\cN(\gY)=0$.  
In particular, if $g$ is Einstein and $(n-8)K\le 0$, with strict 
inequality at some point,
then $\cN(\gS)\cap\cN(\gY)=0$.  
If 
$$
P\le\dfrac{n-8}{8n}K,
$$
with strict inequality at some point, then
$\cN(\gT)\cap\cN(\gZ)=0$.  
In particular, if $g$ is Einstein and $(n-8)K\ge 0$, with strict 
inequality at some point, then
$\cN(\gT)\cap\cN(\gZ)=0$.  
\end{proposition}

\begin{proof}{Proof} By \nnn{bac},
$$
p\le b\le P\;\Rightarrow\; p\le b \,\cdot\le P.
$$
The statements are now immediate from \nnn{elimc}.
\end{proof}

\begin{proposition}\label{dim8} If $n=8$ and $g$ is Einstein, any twistor
in $\cN(\gS)\cap\cN(\gY)$ or $\cN(\gT)\cap\cN(\gZ)$ is parallel.
\end{proposition}

\begin{proof}{Proof} This is immediate from \nnn{elimck}, together
with the fact that $\N=\gS+\gT+\gY+\gZ\,$.
\end{proof}

\section{Mixed Bochner-Weitzenb\"ock formulas}\setcounter{equation}0
In this section, we compute compositions of different gradients which
will play a role in studying necessary conditions for the existence of special
sections of the twistor bundle. Consider the possible compositions
\beq\label{compo}
\bbV(\l)\stackrel{G_{\l\t}}{\longrightarrow}
\bbV(\t)\stackrel{G_{\t\m}}{\longrightarrow}
\bbV(\m),
\eeq
acting between irreducible $\spn$-bundles $\bbV(\l)$ and $\bbV(\m)$
with $\l\ne \m$.
By the selection rule \nnn{selection}, there
are either 0, 1, or 2 compositions \nnn{compo} for a given pair 
$(\l,\m)$.

In fact, the number of compositions \nnn{compo} is
\bdm
r(\l,\m):=\dim\hom_{\spn}(T^*\ot T^*\ot\bbV(\l),\bbV(\m)).
\edm
Since $T^*\ot T^*\cong_{\spn}\tfs^2\os\L^0\os\L^2$, the number
$r(\l,\m)$ breaks up into summands attributable to 
$\tfs^2$, $\L^0$, and $\L^2$:
$$
r(\l,\m)=r_{\tfs^2}(\l,\m)+r_{\L^0}(\l,\m)+r_{\L^2}(\l,\m).
$$
By \cite{socc}, Lemma 2.2,
$$
r(\l,\m)=2\ \implies\ r_{\tfs^2}(\l,\m)=r_{\L^2}(\l,\m)=1.
$$
When $r(\l,\m)=1$, the contribution may come from $\tfs^2$ or 
$\L^2$, depending on the pair $(\l,\m)$.  (See \cite{socc}, Lemma 2.2
for a precise classification.)

Of course, in the complementary 
case $\l=\m$, we have already described these numbers:
$$
r_{\L^0}(\l,\l)=1,\ \ 
r_{\tfs^2}(\l,\l)=[(N(\l)-1)/2],\ \  
r_{\L^2}(\l,\l)=[N(\l)/2].
$$

Since equivariant second-order leading symbols of differential operators
are identified with elements of $\hom_{\spn}(\sym^2\ot\bbV(\l),
\bbV(\m))$, the space of such leading symbols has dimension
$r_{\tfs^2}(\l,\m)$.  (Note that since $\l\ne\m$, we have
$r_{\L^0}(\l,\m)=0$.)
Thus in the case $r(\l,\m)=2$, there is a nontrivial linear
relation
$$
c_1 \,\s_2(G_{\t_1\m}G_{\l\t_1}) + c_2 \,\s_2(G_{\t_2\m}G_{\l\t_2})=0,
$$
where $\t_1\,,\t_2$ are the intermediate weights in \nnn{compo}:
\begin{diagram}
\bbV(\t_1)            ? {} ? \ear[80]
                           {\MV{0}{5}{G_{\t_1\m}}}? {} ? \bbV(\m)             ??
\nar{\MV{5}{0}
{G_{\l\t_1}}}? {} ? {}                ? {} ?\nar{\MV{5}{0}
                                                   {G_{\t_2\m}}}   ?? 
\bbV(\l)            ? {} ? \ear[80]
                           {\MV{0}{5}{G_{\l\t_2}}}? {} ? \bbV(\t_2)              ??
\end{diagram}
As a result, the operator 
$c_1\, G_{\t_1\m}\;G_{\l\t_1}+ c_2\, G_{\t_2\m}\;G_{\l\t_2}$ 
is a curvature action,
and since $r_{\L^0}(\l,\m)=0$, 
the scalar curvature does not contribute to this:
there are actions $\a(C)$ and $\b(b)$ of the Weyl and Einstein
tensors such that
\beq\label{loopprinc}
c_1\, G_{\t_1\m}\;G_{\l\t_1} + c_2\, G_{\t_2\m}\;G_{\l\t_2}=\a(C)+\b(b).
\eeq

When $r(\l,\m)=r_{\L^2}(\l,\m)=1$, there are no second-order
symbols, since then\newline 
$r_{\tfs^2}(\l,\m)=0$.  In addition,
since $b$ is a section
of $\tfs^2$, the Einstein tensor cannot act from $\bbV(\l)$ to
$\bbV(\m)$.  Thus in this case, $G_{\t\m}G_{\l\t}$ is an action
of the Weyl tensor ($\t$ being the unique intermediate weight from
\nnn{compo}); say
\beq\label{lineprinc}
G_{\t\m}\;G_{\l\t}=\a(C).
\eeq

When $\bbV(\l)$ is a twistor bundle, these remarks apply for several
values of $\bbV(\m)$.  Suppose that $n\ge 4$,
and consider the following diagram.

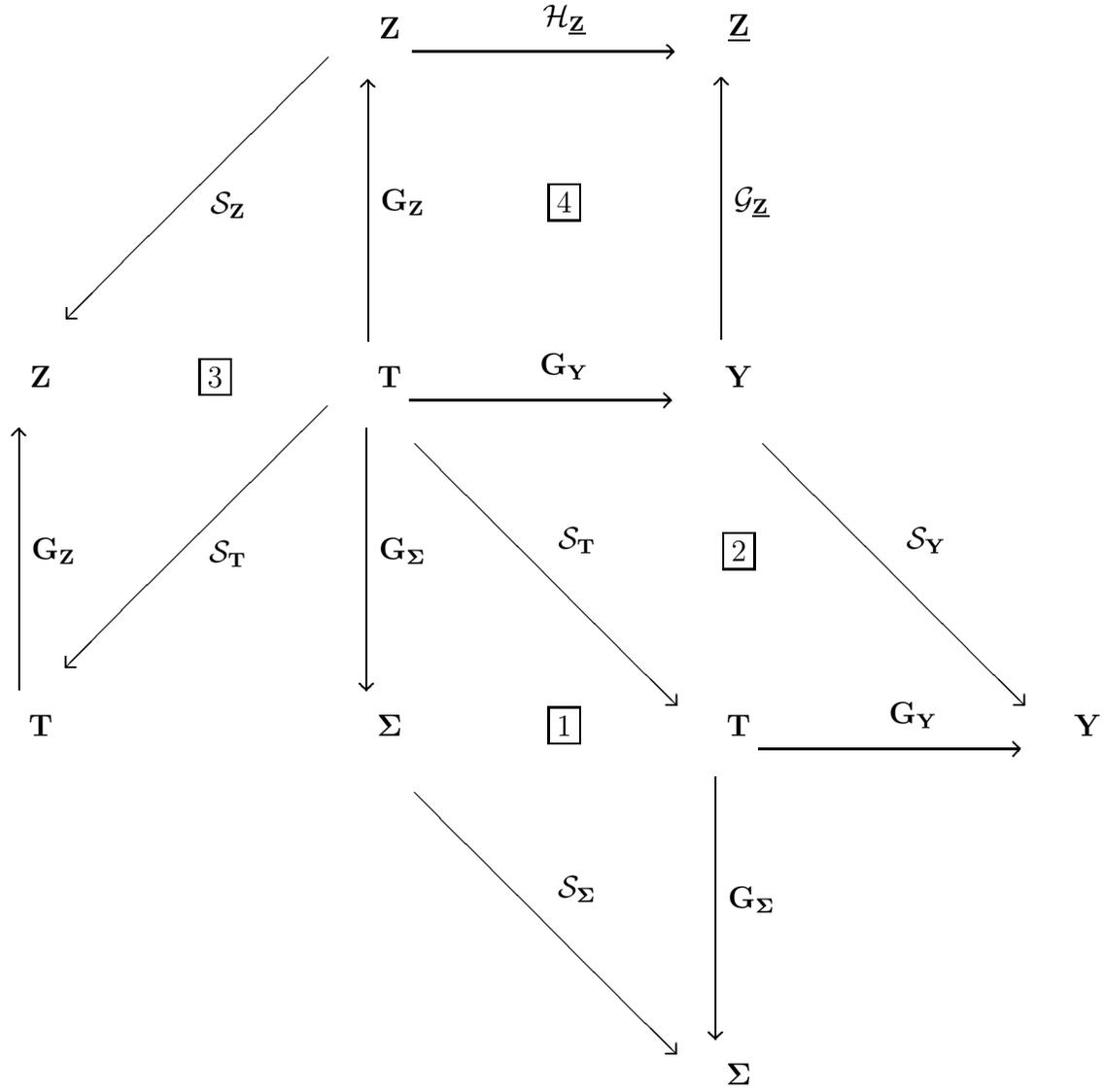
\begin{diagram}[170]\label{mixdiag}
{}     ? {}       ? \Zb    ? \CE\MV05\gZYZ ? \YZb          ??
{}     ? \CSW\MV5{-5}\gZZ 
                  ? \CN\MV50\gZ 
                           ? \fbox{4}      ? \CN\MV50\gYYZ ??
\Zb    ? \fbox{3} ? \Tw    ? \CE\MV05\gY   ? \Yb           ??
\CN\MV50\gZ 
       ? \CSW\MV5{-5}\gTT  
                  ? \CS\MV50\gS 
                           ? \CSE\MV55\gTT  ? \fbox{2} ? \CSE\MV55\gYY ??
\Tw    ? {}       ? \Sp    ? \fbox{1}      ? \Tw      ? \CE\MV05\gY  ?  \Yb ??
{}     ? {}       ? {}     ? \CSE\MV55\gSS  ? \CS\MV50\gS   ??
{}     ? {}       ? {}     ? {}            ? \Sp           ??
\end{diagram}

This contains a new bundle and some new operators, all of which
will be defined and/or computed in the following  sections.
Corresponding to each of the four simple closed loops, we have
an equation of the form \nnn{loopprinc}.  

Strictly speaking, we only have this immediately for $n$ odd, since
each bundle in the diagram is reducible in the even-dimensional case;
they are $\bbU(\l)$ rather than $\bbV(\l)$ bundles.
However, applying the same arguments to their irreducible summands, 
we get the result.  At any rate, we shall compute each relation
explicitly, so we do not really rely on the general principle 
\nnn{loopprinc}.

We also have equations
of the form \nnn{lineprinc} corresponding to the compositions
\beq\label{spinorform}\begin{array}{cc}
\Sp\stackrel{\bT}{\longrightarrow}\Tw
\stackrel{\gY}{\longrightarrow}\Yb\\
\Tw\stackrel{\gY}{\longrightarrow}\Yb
\stackrel{\gYYY}{\longrightarrow}\YYb.
\end{array}
\eeq
\subsection {Mixed BW formulas targeted at spinor-form bundles}\label{mbwsf}

In this subsection we compute the composition \nnn{spinorform} and the 
loops $\fbox{1}$ and $\fbox{2}$.  This brings us into contact
with the theory of {\em spinor-forms} developed in
\cite{tbadv}; see also \cite{soubook}.
The starting point of \cite{tbadv} is the introduction of
variants of 
the exterior and interior operators (both differential
and multiplicative) familiar from the de Rham complex.
On a spinor-$k$-form,
$$
\begin{array}{l}
(\tilde d\f)_{\a_0\,\ldots\a_k}=\sum_{s=0}^k(-1)^s\N_{\a_s}
\f_{\a_0\ldots\hat\a_s\ldots\a_k}\,, \\
(\tilde\d\f)_{\a_2\ldots\a_k}=-\N^\l\f_{\l\a_2\ldots\a_k}\,, \\
(\eg\f)_{\a_0\,\ldots\a_k}=\sum_{s=0}^k(-1)^s\g_{\a_s}
\f_{\a_0\ldots\hat\a_s\ldots\a_k}\,, \\
(\ig\f)_{\a_2\ldots\a_k}=\g^\l\f_{\l\a_2\ldots\a_k}\,.
\end{array}
$$
It is convenient to have a compact notation for the operator 
$$
\bbD=\ig\td+\td\ig=-(\tD\eg+\eg\tD),
$$
which, in index notation, appears as 
$$
(\bbD\f)_{\a_1\,\ldots\,\a_k}=\g^\l\N_\l
\f_{\a_1\,\ldots\,\a_k}\,.
$$
Let $\SF^k$ be the bundle of spinor-$k$-forms. 
The following are identities that can be computed immediately, and which
are used repeatedly:
$$
\begin{array}{rl}
\ig\eg-\eg\ig&=-(n-2k) \\
\ig\tD&=-\tD\ig \\
\eg\td&=-\td\eg \\
\ig\;\bbD+\bbD\;\ig&=2\tD \\
\eg\;\bbD+\bbD\;\eg&=-2\td \,,
\end{array}
$$
and
$$
\begin{array}{rl}
(\td\td\f)_{\a_1\,\ldots\,\a_{k+2}}&=
\sum_{1\le s<t\le k+2}(-1)^{s+t-1} W_{\a_s\a_t}
\f_{\a_1\,\ldots\,\hat{\a_s}\,\ldots\,\hat{\a_t}\,\ldots\,\a_{k+2}}
 \\
((\tD\td+\td\tD-\N^*\N)\f)_{\a_1\,\ldots\,\a_k}&=
-\sum_{s=1}^k(-1)^s\cR^{\a_0}{}_{\a_s}\f_{\a_0\,\ldots\,\hat{\a_s}\,\ldots\,\a_k}
\\
((\tD\;\bbD-\bbD\;\tD)\f)_{\a_2\,\ldots\,\a_k}&=
-\g^\l\cR^{\a_1}{}_\l\f_{\a_1\,\ldots\,\a_k}\\
((\td\;\bbD-\bbD\;\td)\f)_{\a_0\,\ldots\,\a_k}&=
\sum_{s=0}^k(-1)^s\g^\l\cR_{\a_s\l}\f_{\a_0\,\ldots\,\hat{\a_s}\,\ldots\,\a_k}
\\
\bbD^2-\N^*\N&=\tfrac12\g^i\g^j\cR_{ij}\,.
\end{array}
$$
Here $W$ is the spin curvature, and $\cR$ is the ``all-purpose'' curvature,
whose meaning depends on the valence of what sits to its right.  In particular,
in the formula for $\td\td$, it is $W$ rather than $\cR$ that appears -- the 
terms giving the tensorial part of the curvature cancel, for the same
reason that $dd$ vanishes on differential forms.

In particular, for $\y\in\G (\Sp)$, we have $\tD\y=0$ and
\beq\label{id312a}
\begin{array}{c}
(\td\td\y)_{\a\b}= W_{\a\b}\y, \\
(\tD\td+\td\tD-\N^*\N)\y =(\tD\td-\N^*\N)\y = 0,\\
\tD\bbD\y=\bbD\tD\y= 0 , \\
\left((\td\;\bbD-\bbD\;\td)\y\right)_{\a}=\g^\l W_{\a\l} \y = 
- \frac12 r_{\a\l}\g^\l \y\\
\left(\bbD^2-\N^*\N \right) \y=\dfrac{K}{4}\,\y. 
\end{array}
\eeq

For $\f\in\G (\SF^1)$, we get
$$
\begin{array}{rl}
(\td\td\f)_{\a\b\l}&=
W_{\a\b}\f_\l- W_{\a\l}\f_\b+ W_{\b\l}\f_\a , \\
\left((\tD\td+\td\tD-\N^*\N)\f\right)_{\a}&=
\cR^{\b}{}_{\a}\f_{\b}=( W^\b{}_\a+r^\b{}_\a)\f_\b , \\
(\tD\;\bbD-\bbD\;\tD)\f&=-\tfrac12r^\a{}_\b\g^\b\f_\a , \\
\left((\td\;\bbD-\bbD\;\td)\f\right)_{\a_0\a_1}&=
\g^\b\left(-\tfrac12r_{\a_0\b}\f_{\a_1}
-R^\m{}_{\a_1\a_0\b}\f_\m+\tfrac12r_{\a_1\b}\f_{\a_0}
+R^\m{}_{\a_0\a_1\b}\f_\m\right)\,. 
\end{array}
$$

Let $\SF^k_{{\rm top}}$ be the subbundle of $\SF^k$ 
annihilated by $\i(\g)$.
For example, $\SF^0_{{\rm top}}=\Sp$,
$\SF^1_{{\rm top}}=\Tw$, $\SF^2_{{\rm top}}=\Yb$.  The bundle
$\YYb$ from \nnn{spinorform} can now be defined as $\SF^3_{{\rm top}}$.
For $(n-2k)(n-2k+1)\ne 0$, when we compress $\td$ to an operator
going from $\SFtop^k$ to $\SFtop^{k+1}$,
we get
\beq\label{edtop}
\tdtop_k=\td+\frac1{n-2k}\eg\;\bbD+\frac1{(n-2k)(n-2k+1)}\eg^2\tD\,.
\eeq
For $k=0,1,2$, these operators are related to those used in diagram 
\nnn{mixdiag} by
\beq\label{id312b}
\tdtop_0 = \bT = \frac12 \;\agS I_{{\bf\S}}, \qquad \tdtop_1 = 2\gY, 
\qquad 
\tdtop_2 := \gYYY .
\eeq
On weight theoretic grounds, we can predict that $\tdtop_{k+1}\;\tdtop_k$
is an action of the Weyl tensor, at least provided we stay safely below
the middle order of form.
Indeed, using the above identities, we readily obtain the result that
$\tdtop_{k+1}\;\tdtop_k$ is a curvature action.
But by \nnn{lineprinc}, there is no action of symmetric two-tensors carrying
$\SFtop^k$ to $\SFtop^{k+2}$; thus the Ricci tensor cannot act.
Thus the action depends the Riemann curvature only through the
Weyl tensor.  For $k=0$ or $k=1$, the above identities yield to the following 
Weyl curvature actions
\beq\label{weylaction}
\begin{array}{rl}
(\tdtop_1\;\tdtop_0\y)_{\a\b}&=-\dfrac14C_{\l\m\a\b}\g^\l\g^\m\y, \nonumber\\
(\tdtop_2\;\tdtop_1\f)_{\a_0\a_1\a_2}&=6\left\{
-\dfrac18C_{\l\m[\a_0\a_1}\g^\l\g^\m\f_{\a_2]}
+\dfrac1{n-4}C^\l{}_{[\a_2\a_1\m}\g_{\a_0]}\g^\m\f_\l\right. \nonumber\\
&\qquad\left.
+\dfrac1{4(n-4)(n-3)}C_{\l\m}{}^\b{}_{[\a_2}\g_{\a_0}\g_{\a_1]}\g^\l\g^\m\f_\b
\right\} \\
&:=(\a(C)\f)_{\a_0\a_1\a_2}\,,
\end{array}
\eeq
where square brackets denote antisymmetrization.  (For the second
identity,
assume $n>4$.)
The extreme right-hand side is in fact (up to constant multiples) the
unique action of the Weyl tensor from $\SFtop^1$ to $\SFtop^3$.
In particular, if we take any of its three terms, each of which
is valued in $\SF^3$, and project to $\SFtop^3$, the whole expression
will emerge. With the notation of diagram \nnn{mixdiag}, 
the above relations translate to

\begin{proposition}\label{ddiff}
For $n\ge 4$, the compositions \nnn{spinorform} are given by
\beq
\begin{array}{rl}
(\gY \;\bT \y)_{\a\b} &= - \frac18\;C_{\l\m\a\b}\g^\l\g^\m\y,\\
\gYYY \;\gY \f &= \frac12\;\a(C)\f \,,
\end{array}
\eeq
where the action of the Weyl tensor $\a(C)$ is given by \nnn{weylaction}.
\end{proposition}

We are also interested in the self-gradient on $\SFtop^k$, since
the case $k=2$ enters in our considerations.
We can get this by compressing the conformally covariant
operator $P_k$ of \cite{tbadv} from $\SF^k$ to 
$\SFtop^k$.  Since
$$
P_k=\dfrac{n-2k+4}2\;\ig\td + \dfrac{n-2k}2\;\Big(\td\ig-\tD\eg\Big)
-\dfrac{n-2k-4}2\;\eg\tD,
$$
we have
\beq\label{esgsf}
\begin{array}{rl}
P_k|_{\SFtop^k}&=\left(\frac{n-2k+4}2\;\bbD-\frac{n-2k}2\;\tD\eg
-\frac{n-2k-4}2\;\eg\tD\right)\bigg|_{\SFtop^k} \\
&=\left((n-2k+2)\;\bbD+2\eg\tD\right)\bigg|_{\SFtop^k}\,,
\end{array}
\eeq
since $\ig$ annihilates $\SFtop^k$.  Some weight theory (including
the conformal weights of the operators involved) predicts that the
restriction of $P_k$ to 
$\SFtop^k$ will also be the compression; that is, that $P_k$ carries
$\SFtop^k$ to itself.  We can check this by computing
\beq\label{iggypop}
\ig\left((n-2k+2)\;\bbD+2\eg\tD\right)\bigg|_{\SFtop^k}\,;
\eeq
this should vanish.
Using our list of identities to move $\ig$ to the
right, we get
$$
\begin{array}{rl}
\ig\;\bbD\bigg|_{\SFtop^k}&=2\tD\bigg|_{\SFtop^k}\,, \\
\ig\eg\tD\bigg|_{\SFtop^k}&=-(n-2k+2)\tD\bigg|_{\SFtop^k}\,,
\end{array}
$$
so \nnn{iggypop} vanishes as predicted.

Let $\tilde\cS_{k}=P_k|_{\SFtop^k}\,$.  Then, for $k=0$ in \nnn{esgsf}, one 
gets
\beq\label{sgsf0}
\tilde\cS_{0} = (n+2)\; \dc .
\eeq
For $k=1$, and for any $\f\in \G (\Tw)$, we obtain
\beq
\begin{array}{l}
(\tilde\cS_{1} \;\f)_\a = \left((n\;\bbD + 2\eg\tD)\f\right)_\a
= n \;\g^\l\N_\l \f_\a - 2 \g_\a \N^\l\f_\l\,.
\end{array}
\eeq
Thus by \nnn{rar1}, 
\beq\label{sgsf1}
\tilde\cS_{1} = n\rso.
\eeq

Letting 
$k=2$ in \nnn{esgsf}, we 
{\em denote} the self-gradient $\tilde\cS_{2}$ by $\gYY$. 
For any $\f\in\G (\Yb )$, we get 
\beq\label{sgsf2}
\begin{array}{l}
(\gYY \,\f)_{\a\b} = (n-2)\;\g^\l\N_\l \f_{\a\b}  
-2\left(\g_\a\N^\l\f_{\l\b} - \g_\b\N^\l\f_{\l\a}\right)\,.
\end{array}
\eeq
By \nnn{edtop} and \nnn{esgsf}, there   
is a linear relation between the leading symbols 
of $\tilde\cS_{k+1}\;\tdtop_k$ and $\tdtop_k\;\tilde\cS_{k}\,$.  
We may calculate this directly as follows:
let $s=n-2k$, and let ``$\sim$'' be equality modulo a curvature action.
Then
\beq\label{ledboots}
\begin{array}{ll}
\tilde\cS_{k+1}\;\tdtop_k &=(s-2)\td\;\bbD-\dfrac{s+2}{s}\eg\;\bbD^2
-\dfrac{s-2}{s(s+1)}\eg^2\tD\;\bbD+\dfrac4{s}\eg\td\tD+2\eg\tD\td, \\
\tdtop_k\;\tilde\cS_{k} &=(s+2)\td\;\bbD+\dfrac{s+2}{s}\eg\;\bbD^2
-\dfrac{s+2}{s(s+1)}\eg^2\tD\;\bbD
-\dfrac{2(s+2)}{s}\eg\td\tD.
\end{array}
\eeq
As a result, 
$$
\tilde\cS_{k+1}\;\tdtop_k-\frac{s-2}{s+2}\;\tdtop_k\;\tilde\cS_{k}\sim
2\eg\;\Big(-\bbD^2+\tD\td+\td\tD\Big)\sim 0;
$$
this is the desired linear relation.

Keeping track of lower-order terms, we use the above identities
to compute that
\beq\label{eq312a}
\begin{array}{l}
\tilde\cS_{k+1}\,\tdtop_k-\dfrac{s-2}{s+2}\,\tdtop_k\,\tilde\cS_{k}
=s\;[\bbD,\td]+\dfrac3{s+1}\;\eg^2\;[\bbD,\tD] \\
{}\qquad+
\dfrac4{(s+1)(s+2)}\;\eg^3\;\tD\tD+2\eg\Big(\tD\td+\td\tD-\bbD^2\Big).
\end{array}
\eeq
In particular, for $k=0$, using
\nnn{id312a} to compute the 
right side of \nnn{eq312a}, we get
\beq\label{mBWf1}
\tilde\cS_{1}\,\tdtop_0-\frac{n-2}{n+2}\,\tdtop_0\;\tilde\cS_{0}=
\frac{n}{2}\, {\bf b}_{{\bf T}\to{\bf\S}}(b)^* ,
\eeq
where the action of the Einstein tensor $b$ taking a twistor $\f$ to a spinor
is given by
\bdm
{\bf b}_{{\bf T}\to{\bf\S}}(b) \f := -b_{\a\b} \g^\a \f^\b ,
\edm
and its adjoint, taking a spinor field $\y$ to
a twistor, is 
\beq\label{aeloop1}
({\bf b}_{{\bf T}\to{\bf\S}}(b)^*\, \y )_\a := b_{\a\b} \g^\b \y.
\eeq
For $k=1$, we get
\beq\label{mBWf2}
\tilde\cS_{2}\;\tdtop_1-\frac{n-4}{n}\;\tdtop_1\;\tilde\cS_{1} =
{\bf C}(C) +(n-4) {\bf b}(b),
\eeq
where 
\beq\label{wloop2}
({\bf C}(C)\f)_{\a\b} := (n-2)\; C^\l{}_{\m\a\b} \g^\m \f_\l
-C_{\l\m}{}^\nu{}_{[\a}\g_{\b]} \g^\l \g^\m \f_\nu
\eeq
and
\beq\label{eloop2}
({\bf b}(b)\f)_{\a\b}:=\g^\l b_{\l[\a}\f_{\b]}
-\frac2{n-2}\;b^\l{}_{[\a}\g_{\b]}\f_\l
+\frac1{(n-1)(n-2)}\;\g_{\a\b}b^\l{}_\m\g^\m\f_\l\,.
\eeq
Note that ${\bf b}$ is the unique action of the Einstein tensor
carrying $\SFtop^1$ to $\SFtop^2\,$.
Up to a constant factor, it could already have been predicted by
looking at either operator in \nnn{ledboots}:
the leading symbol of each must exhibit the unique $\tfs^2$ action:
replacing $b$ by $\xi\otimes\xi-|\xi|^2g/n$ in \nnn{eloop2}
produces a constant multiple of the leading symbol, evaluated at the
covector
$\xi$.

With the identifications \nnn{id312b}, 
\nnn{sgsf0}, \nnn{sgsf1} and \nnn{sgsf2}, formulas \nnn{mBWf1} and 
\nnn{mBWf2} yield
\begin{proposition}\label{wang1}
The identities of \nnn{loopprinc} corresponding
to the adjoint of loop $\fbox{{\rm 1}}$ and to loop $\fbox{{\rm 2}}$ 
are realized by
\bdm
\begin{array}{l}
\rso\; \bT - \frac{n-2}{n}\; \bT\; \dc = \frac12\; 
{\bf b}_{{\bf T}\to{\bf\S}}(b)^* 
\qquad {\rm for}
\quad n\ge 2 ,\\
2\gYY\;\gY-2(n-4)\gY\,\rso =
{\bf C}(C)+(n-4) {\bf b}(C)\qquad {\rm for}
\quad n\ge 4,
\end{array}
\edm
where the curvature actions are given by \nnn{aeloop1}, \nnn{wloop2} and 
\nnn{eloop2}.
\end{proposition}
Note that the first formula in Proposition \ref{wang1} 
is proved in \cite{wang}.  The second formula, in the case $n=4$, is
actually an additional realization of the second formula in Proposition
\ref{ddiff} above: their abstract targets are realized within both 
spinor-3-forms and spinor-2-forms.

\subsection {Mixed BW formulas targeted at other tensor-spinor 
bundles}\label{mbwts}

In this section, we compute the instance of formula \nnn{loopprinc}
corresponding to  
loop $\fbox{3}$. For this, we need to compute
the self-gradient
$\gZZ$ that acts in the target bundle $\Zb$ for $\gZ$.  This target consists
of spinor-2-tensors $\f=(\f_{\a\b})$ which are trace free and symmetric
in the two tensor arguments, and which are annihilated by interior
Clifford multiplication in the sense that $\g^\a\f_{\a\b}=0$.  
As in the remark after \nnn{annih}, the trace-free condition is actually
redundant.

It is 
not difficult to compute that the operator
\beq\label{selfZ}
(\gZZ^0\f)_{\a\b}=(\bbD\f)_{\a\b}-\frac2{n+2}(\g_\a\N^\l\f_{\l\b}+
\g_\b\N^\l\f_{\l\a})
\eeq
has its range in $\Zb$.  Being manifestly equivariant, 
it must be a realization of the self-gradient
if $n$ is odd, and of the gradients 
$$
\bbV(\tfrac52,\tfrac12,\ldots,\tfrac12,\pm\tfrac12)\to
\bbV(\tfrac52,\tfrac12,\ldots,\tfrac12,\mp\tfrac12)
$$
for $n$ even.
Here and below, just as in the spinor-form case, 
$\bbD$ is the Dirac expression $\g^\l\N_\l\,$, which
can act on any bundle $\Sp\ot\bbT$ for which $\bbT$ is a tensor bundle.
This particular normalization of the operator has no special meaning,
but it seems convenient to have coefficient 1 on the $\bbD$ part.
Recall that we adopted the same convention to normalize the Rarita-Schwinger
operator $\rso$:
$$
(\rso\f)_\b=(\bbD\;\f)_\b-\frac2{n}\g_\b\;\dv\f,
$$
where 
$$
\dv\;\f=\N^\l\f_\l\,.
$$
Using these expressions and the 
explicit expression \nnn{egZ} for $\gZ$, we find
a relation between $\gZZ^0\,\gZ$ and $\gZ\,\rso$ on the leading symbol level:
\bdm
\s_2(\gZZ^0\,\gZ) = \frac{n}{n+2}\;\s_2(\gZ\;\rso).
\edm
In fact, each side just above has the same second-order symbol as 
\beq\label{dd5}
\begin{array}{ll}
&\dfrac12\Big(\N_\a(\bbD\,\f)_\b + \N_\b(\bbD\,\f)_\a\Big) 
-\dfrac1{n+2}\Big(\g_\b\N_\a\dv\,\f + \g_\a\N_\b\dv\,\f\Big) \\
&\qquad+\dfrac1{2(n+2)}\Big(\g_\a(\bbD^2\,\f)_\b + \g_\b(\bbD^2\,\f)_\a\Big) 
-\dfrac1{n+2}\,g_{\a\b}\,\bbD\,\dv\,\f.
\end{array}
\eeq
That there should be such a relation between leading symbols is
expected,
by \nnn{loopprinc}.

The difference
\beq\label{diff52}
\frac{n}{n+2}\,\gZ\,\rso - \gZZ^0\,\gZ
\eeq
is some curvature action $\Tw\to\Zb$, and computing a little more, we
can find it.  Note that this curvature action cannot involve the scalar
curvature (since $\Zb\not\cong_{\spin(n)}\Tw$).  
Up to constant multiples, there
is just one action of the Einstein tensor which can appear, since
$\tfs^2\ot\Tw$ contains just one copy
of $\Zb$; this is the same fact used to obtain \nnn{loopprinc}.
This action must already be visible in the leading symbol of the
operator \nnn{dd5}: replacing each $\N\N$ in this formula by $b$
(noting that $\N$ is implicit in $\bbD$ and div), we get $\frac12
\underline{{\bf b}}(b)$, where
\beq\label{bb52}
(\underline{{\bf b}}(b)\f)_{\a\b}=\g^\m(b_{\a\m}\f_\b+b_{\b\m}\f_\a)
-\frac2{n+2}\left\{b_\a{}^\l\g_\b+b_\b{}^\l\g_\a+g_{\a\b}b^\l{}_\m\g^\m
\right\}\f_\l\,.
\eeq

It is not immediately clear how many Weyl tensor actions
carry $\Tw$ to $\Zb$, but in fact, trying all the combinatorial 
possibilities, it is straightforward to show that there is just one:
\beq\label{cc52}
(\underline{{\bf C}}(C)\f)_{\a\b}=\g^\l\Big(C^\nu{}_{\b\l\a}
+C^\nu{}_{\a\l\b}\Big)\f_\nu
-\frac3{2(n+2)}\Big(\g_\a C^\l{}_{\b\k\nu}+\g_\b C^\l{}_{\a\k\nu}\Big)
\g^\k\g^\nu\f_\l\,.
\eeq

Computing the difference \nnn{diff52} explicitly, one obtains:

\begin{proposition} A realization of 
\nnn{loopprinc} for loop $\fbox{3}$ in diagram \nnn{mixdiag} is given by
$$
\frac{n}{n+2}\,\gZ\rso - \gZZ\gZ = -\frac{n}{4(n-2)}\,\underline{{\bf b}}
(b) 
+ \frac12 \,\underline{{\bf C}}(C),
$$
where the Einstein and Weyl actions are given by
\nnn{bb52} and \nnn{cc52}.
\end{proposition}

\subsection {Mixed BW formulas with target in 
higher tensor-spinor bundles}\label{mbwhts}

The new objects in diagram \nnn{mixdiag} and \nnn{spinorform}
are defined as follows.
The bundle $\YZb$ is a tensor-spinor realization of 
$$
\bbU(\tfrac52\,,\tfrac32\,,\tfrac12\,,\ldots,\tfrac12).
$$
There are two competing realizations of this bundle, of approximately
the same complexity.
To describe these, it is convenient to first describe a corresponding
pair of competing realizations of $\bbU(2,1,0,\ldots,0)$.
As always, the tensor realizations and differential operator
formulas speak for themselves, and 
an understanding of the representation-theoretic background is not
strictly required.

Consider tensors $\f_{\l\a\b}$ in $T^*\ot\L^2$; that is, tensors
with the symmetry $\f_{\l\a\b}=-\f_{\l\b\a}\,$.  Under the action
of $\orth(n)$, there are three projections of such tensors.  The
$\L^3$ part is
$$
(P_{\L^3}\f)_{\l\a\b}=\dfrac13\left(\f_{\l\a\b}+\f_{\a\b\l}+\f_{\b\l\a}\right).
$$
The remaining parts, being orthogonal to this, must satisfy the 
Bianchi-like identity
\beq\label{bianchilike}
\k_{\l\a\b}+\k_{\a\b\l}+\k_{\b\l\a}=0.
\eeq
The $\L^1$ part is
$$
(P_{\L^1}\f)_{\l\a\b}=\dfrac1{n-1}
\left(g_{\l\a}\f^\m{}_{\m\b}+g_{\l\b}\f^\m{}_{\a\m}\right).
$$
(Up to a constant multiple, this is the only ``pure trace''
that is antisymmetric in the second and third arguments.  The
constant $1/(n-1)$ is determined by the projection condition.)
Note that $\k=P_{\L^1}\f$ satisfies the Bianchi-like identity
\nnn{bianchilike}.
The remaining part is
$$
(P\f)_{\l\a\b}=\dfrac23\f_{\l\a\b}-\dfrac13\f_{\a\b\l}
-\dfrac13\f_{\b\l\a}
-\dfrac1{n-1}\left(g_{\l\a}\f^\m{}_{\m\b}+g_{\l\b}\f^\m{}_{\a\m}\right).
$$
One may check that 
$$
(P\f)^\a{}_{\a\b}=0,
$$
and that 
$\k=P\f$ satisfies \nnn{bianchilike}.
The bundle we have reached must be isomorphic 
to $\bbV(2,1,0,\ldots,0)$ by the selection rule \nnn{selection}; its
symmetry type is: (1) antisymmetric
in the last two arguments; (2) totally trace-free; (3) Bianchi-like
in the full three arguments.

Now consider a tensor in $\y\in T^*\ot\tfs^2$.
The projection onto the symmetric $3$-tensors $\sym^3$ is
$$
\dfrac13\left(\y_{\l\a\b}+\y_{\a\b\l}+\y_{\b\l\a}\right).
$$
But $\sym^3$ splits under O$(n)$, into the direct sum of $\tfs^3$ 
and $\L^1=\tfs^1$.  The projection of $\y$ onto
$\tfs^3$ will take the form
$$
(Q_{\tfs^3}\y)_{\l\a\b}=\dfrac13\left(\y_{\l\a\b}+\y_{\a\b\l}+
\y_{\b\l\a}\right)
-a\left(g_{\b\l}\y^\m{}_{\m\a}+g_{\a\l}\y^\m{}_{\m\b}
+g_{\a\b}\y^\m{}_{\m\l}\right),
$$
where $a$ is some constant.  The requirement that the $\a\b$-trace
(and thus all traces) vanish gives 
$a=\dfrac2{3(n+2)}\,$:
$$
(Q_{\tfs^3}\y)_{\l\a\b}=\dfrac13\left(\y_{\l\a\b}+\y_{\a\b\l}
+\y_{\b\l\a}\right)
-\dfrac2{3(n+2)}
\left(g_{\b\l}\y^\m{}_{\m\a}+g_{\a\l}\y^\m{}_{\m\b}
+g_{\a\b}\y^\m{}_{\m\l}\right).
$$

The $\L^1$ projection
will have the form
\beq\label{theproj}
(Q_{\L^1}\y)_{\l\a\b}=
c_1\left(g_{\l\a}\y^\m{}_{\m\b}+g_{\l\b}\y^\m{}_{\a\m}\right)
+c_2g_{\a\b}\y^\m{}_{\m\l}\,,
\eeq
where $c_1$ and $c_2$ are constants.
The projection condition leads to the system
\bdm
\begin{array}{l}
c_2=c_2\{c_2+(n+1)c_1\}, \\
c_1=c_1\{c_2+(n+1)c_1\}.
\end{array}
\edm
Thus the projection is trivial unless
\begin{equation}\label{eqn1}
c_2+(n+1)c_1=1.
\end{equation}
The trace-free condition in $\a\b$ gives
$$
2c_1+nc_2=0, 
$$
and the last two equations force
$$
c_2=-\dfrac2{(n+2)(n-1)}\,,\qquad c_1=\dfrac{n}{(n+2)(n-1)}\,.
$$
(Note that (\ref{eqn1}) implies that the $\l\a$ trace of 
(\ref{theproj}) is $\y^\m{}_{\m\b}\,$.)
Collecting this information, we have
$$
(Q_{\L^1}\y)_{\l\a\b}=
\dfrac{n}{(n+2)(n-1)}\Big(g_{\l\a}\y^\m{}_{\m\b} +
g_{\l\b}\y^\m{}_{\a\m}\Big)
-\dfrac2{(n+2)(n-1)}g_{\a\b}\y^\m{}_{\m\l}\,.
$$

The remaining projection is
\bdm
\begin{array}{rl}
(Q\y)_{\l\a\b} &=
\dfrac23\y_{\l\a\b}-\dfrac13\y_{\a\b\l}
-\dfrac13\y_{\b\l\a} + \dfrac2{3(n-1)}g_{\a\b}\y^\m{}_{\m\l}\\
&\qquad -\dfrac1{3(n-1)}\Big(g_{\l\a}\y^\m{}_{\m\b}
+ g_{\l\b}\y^\m{}_{\a\m}\Big)
\,.
\end{array}
\edm
Note that $\k=Q\y$ satisfies \nnn{bianchilike}, and that
$(Q\y)^\a{}_{\a\b}=0$.
By the selection rule \nnn{selection}, we have landed in a copy of
$\bbV(2,1,0,\ldots,0)$, and by the above, we have 
landed in the following symmetry type:
(1) symmetric
in the last two arguments; (2) totally trace-free; (3) Bianchi-like
in the full three arguments.

The isometry between the two competing realizations of
$\bbV(2,1,0,\ldots,0)$ is
\beq\label{isomes}
\begin{array}{l}
\y'_{\l\a\b}=-\dfrac1{\sqrt{3}}(\f_{\a\b\l}+\f_{\b\a\l}), \\
\f'_{\l\a\b}=\dfrac1{\sqrt{3}}(\y_{\a\b\l}-\y_{\b\a\l}).
\end{array}
\eeq
That is, denoting the two tensor bundles by $\bbV_P$ and $\bbV_Q\,$,
the maps
$$
\begin{array}{c}
\bbV_P\leftrightarrow\bbV_Q\,, \\
\f\mapsto\y', \\
\f'\leftarrow\y
\end{array}
$$
are isometries.
One could also reverse the roles of $\pm 1/\sqrt{3}$ in \nnn{isomes}.

This material on $\bbV(2,1)$ is significant because
$\bbV(\frac52\,,\frac32)$ is the Cartan product (highest
weight direct summand) of $\Sp\ot\bbV(2,1)$.  We may thus
realize $\bbV(\frac52\,,\frac32)$ as the bundle of tensor-spinors
in $\Sp\ot\bbV_P\,$, or in $\Sp\ot\bbV_Q\,$, satisfying
the interior multiplication conditions
\beq\label{interiors}
\g^\l\k_{\l\a\b}=0,\ \g^\a\k_{\l\a\b}=0.
\eeq
We denote by $(\Sp\ot\bbV_P)_{{\rm top}}$ and $(\Sp\ot\bbV_Q)_{{\rm top}}$
the subbundles cut out by this condition.
This allows us to compute realizations of the gradients
$$
\bbU(\tfrac32\,,\tfrac32)\to
\bbU(\tfrac52\,,\tfrac32)\ \ {\rm and}\ \ \bbU(\tfrac52)\to
\bbU(\tfrac52\,,\tfrac32)
$$
(where, for convenience, we have omitted terminal strings of $\tfrac12$'s)
as differential operators carrying
$$
\begin{array}{l}
\Yb\to
(\Sp\ot\bbV_P)_{{\rm top}}\,, \\
\Zb\to(\Sp\ot\bbV_Q)_{{\rm top}}\to(\Sp\ot\bbV_P)_{{\rm top}}\,,
\end{array}
$$
the very last arrow by the isometry between $\bbV_P$ and $\bbV_Q\,$.
That is, we agree on one of the competing realizations for 
$\bbV(\frac52,\frac32)$, namely
$$
\YZb:=(\Sp\ot\bbV_P)_{{\rm top}}\,,
$$
for purposes of comparing the operators $\gYYZ$ and $\gZYZ$ in loop $\fbox{4}$
of diagram \nnn{mixdiag}.

To compute the gradient $\gYYZ$,
we first compute the projection $\Pi$ of 
$T^*\ot\Yb$
onto $\YZb$.  We will then have
$$
\gYYZ\;\eta=\Pi(\nabla\eta).
$$ 
$\Pi\f$ should have the form
\bdm
\begin{array}{rl}
(\Pi\f)_{\l\a\b}&=\frac23\f_{\l\a\b}-\frac13\f_{\a\b\l}-\frac13\f_{\b\l\a} \\
&\quad +a_1\g_\l\g^\m\f_{\m\a\b}+a_2\left(\g_\a\g^\m\f_{\m\l\b}-\g_\b\g^\m\f_{\m\l\a}\right) \\
&\quad +a_3\left(g_{\l\a}\f^\m{}_{\m\b}-g_{\l\b}\f^\m{}_{\m\a}\right)
+a_4\left(\g_\l\g_\a\f^\m{}_{\m\b}-\g_\l\g_\b\f^\m{}_{\m\a}\right) \\
&\quad +a_5\left(\g_\a\g_\b-\g_\b\g_\a\right)\f^\m{}_{\m\l}
\end{array}
\edm
for some constants $a_i\,$.

We now impose the interior multiplication conditions \nnn{interiors};
after some calculation, we obtain
$$
\begin{array}{l}
a_1=\dfrac2{3(n+2)}\,,\ 
a_2=\dfrac1{3(n+2)}\,,\ 
a_3=-\dfrac{3n+4}{3n(n+2)}\,,\\ 
a_4=-\dfrac1{3n(n+2)}\,,\ 
a_5=\dfrac1{3n(n+2)}\,.
\end{array}
$$
(Note, in this connection, 
that the trace-free conditions actually follow from the interior
multiplication conditions and the Clifford relations, as in the
remark after \nnn{annih}.)
The Bianchi-like identity \nnn{bianchilike} now holds automatically
for $\k=\Pi\f$; in fact, 
this just depends on the conditions
$$
a_1-2a_2=a_4+a_5=0.
$$

To get the operator $\gZYZ$, we first compute the projection onto
$(\Sp\ot\Zb)_{{\rm top}}$ of $\y\in T^*\ot\Zb$; this must have the
form
\bdm
\begin{array}{rl}
(\Xi\y)_{\l\a\b}&=\frac23\y_{\l\a\b}-\frac13\y_{\a\b\l}-\frac13\y_{\b\l\a} \\
&\quad +b_1\g_\l\g^\m\y_{\m\a\b}+b_2\left(\g_\a\g^\m\y_{\m\l\b}+
\g_\b\g^\m\y_{\m\l\a}\right) \\
&\quad +b_3\left(g_{\l\a}\y^\m{}_{\m\b}+g_{\l\b}\y^\m{}_{\m\a}\right) 
+b_4g_{\a\b}\y^\m{}_{\m\l} \\
&\quad +b_5\left(\g_\l\g_\a\y^\m{}_{\m\b}+\g_\l\g_\b\y^\m{}_{\m\a}\right)
\end{array}
\edm
for some constants $b_i\,$.  The interior multiplication conditions
give, after some calculation,
$$
b_1=\dfrac2{3(n-2)}\,,\ b_2=b_3=-\dfrac1{3(n-2)}\,,\ 
b_4=\dfrac{2(n-3)}{3n(n-2)}\,,\ b_5=-\dfrac1{n(n-2)}\,.
$$
$\Xi\y$ then automatically satisfies \nnn{bianchilike};
this just
depends on the relations
$$
b_1+2b_2=2b_3+b_4-2b_5=0.
$$
To reach the $\YZb$ realization, we now apply the isometry
\nnn{isomes} in the tensorial factor:
\bdm
\begin{array}{rl}
\sqrt{3}(\tilde\Xi\y)_{\l\a\b}
&=(\Xi\y)_{\a\b\l}-(\Xi\y)_{\b\a\l} \\
&=\y_{\a\b\l}-\y_{\b\l\a}
+{\dfrac1{n-2}}
\Big(\g_\a\g^\m\y_{\m\b\l}-\g_\b\g^\m\y_{\m\a\l}\Big) \\
&\quad -\dfrac1{n}
\Big(g_{\a\l}\y^\m{}_{\m\b}-g_{\b\l}\y^\m{}_{\m\a}\Big) \\
&\quad -\dfrac1{n(n-2)}
\Big\{(\g_\a\g_\b-\g_\b\g_\a)\y^\m{}_{\m\l}
+\g_\a\g_\l\y^\m{}_{\m\b}-\g_\b\g_\l\y^\m{}_{\m\a}\Big\}.
\end{array}
\edm

By the above, our two gradient realizations are
\bdm
\begin{array}{rl}
(\gYYZ\;\f)_{\l\a\b}&:=\dfrac23\nd_\l\f_{\a\b}-\dfrac13\nd_\a\f_{\b\l}
-\dfrac13\nd_\b\f_{\l\a} \\
&\quad +\dfrac2{3(n+2)}\g_\l\g^\m\nd_\m\f_{\a\b}+\dfrac1{3(n+2)}
\Big(\g_\a\g^\m\nd_\m\f_{\l\b}
-\g_\b\g^\m\nd_\m\f_{\l\a}\Big) \\
&\quad -\dfrac{3n+4}{3n(n+2)}(g_{\l\a}\nd^\m\f_{\m\b}
-g_{\l\b}\nd^\m\f_{\m\a})\\
&\quad -\dfrac1{3n(n+2)}\Big(\g_\l\g_\a\nd^\m\f_{\m\b}-\g_\l\g_\b\nd^\m\f_{\m\a}\Big) \\
&\quad +\dfrac1{3n(n+2)}\Big(\g_\a\g_\b-\g_\b\g_\a\Big)\nd^\m\f_{\m\l}\,,
\end{array}
\edm
and $1/\sqrt{3}$ times
\bdm
\begin{array}{rl}
(\gZYZ\;\y)_{\l\a\b}&:=
\nd_\a\y_{\b\l}-\nd_\b\y_{\l\a}
+\dfrac1{n-2}
\Big(\g_\a\g^\m\nd_\m\y_{\b\l}-\g_\b\g^\m\nd_\m\y_{\a\l}\Big) \\
&\quad -\dfrac1{n}
\Big(g_{\a\l}\nd^\m\y_{\m\b}-g_{\b\l}\nd^\m\y_{\m\a}\Big) \\
&\quad -\dfrac1{n(n-2)}
\Big\{(\g_\a\g_\b-\g_\b\g_\a)\nd^\m\y_{\m\l}
+\g_\a\g_\l\nd^\m\y_{\m\b}-\g_\b\g_\l\nd^\m\y_{\m\a}\Big\}.
\end{array}
\edm

What will be important are the compositions
$$
\gYYZ \;\gY\ \ {\rm and}\ \ \gZYZ \;\gZ\,,
$$
each of which carries $\Tw$ to $\YZb\,$.
The principal part of each composition turns out to be the following: it
carries a section $\F$ of $\Tw$ to
\bdm
\begin{array}{l}
-\dfrac{n^2-2n-2}{2(n+2)(n-2)}\left\{\F_{\a|\b\l}-\F_{\b|\a\l}\right\} \\
+\dfrac{n^2-n-3}{2n(n+2)(n-2)}\left\{g_{\b\l}\F_{\a|\m}{}^\m
                                   -g_{\a\l}\F_{\b|\m}{}^\m\right\} \\
+\dfrac1{2(n+2)}\left\{\g_{\m\l}(\F_{\a|\b}{}^\m-\F_{\b|\a}{}^\m)
                     -g_{\b\l}\F_{\m|\a}{}^\m+g_{\a\l}\F_{\m|\b}{}^\m\right\}\\
+\dfrac1{2n(n+2)(n-2)}\left\{\g_{\b\l}\F_{\a|\m}{}^\m-\g_{\a\l}\F_{\b|\m}{}^\m
                      \right\} \\
-\dfrac{n}{2(n+2)(n-2)}\left\{\g_{\b\m}\F_{\a|}{}^\m{}_\l
                            -\g_{\a\m}\F_{\b|}{}^\m{}_\l\right\} \\
+\dfrac1{2(n+2)(n-2)}\left\{\g_{\b\l}\F_{\m|\a}{}^\m-\g_{\a\l}\F_{\m|\b}{}^\m
                           +g_{\a\l}\g_{\b\m}\F_{\nu|}{}^{\m\nu}
                           -g_{\b\l}\g_{\a\m}\F_{\nu|}{}^{\m\nu}\right\} \\
-\dfrac1{(n+2)(n-2)}\left\{\g_{\a\b}\F_{\m|}{}^\m{}_\l
                          +\g_{\b\m}\F_{\l|\a}{}^\m
                          -\g_{\a\m}\F_{\l|\b}{}^\m\right\} \\
-\dfrac1{n(n+2)(n-2)}\g_{\a\b}\F_{\l|\m}{}^\m ,
\end{array}
\edm

where we have used the notation
$$
\F_{\a|\b\l} : = \nd_\l\nd_\b \F_\a.
$$
This computation tells us, as a bonus, 
what the single (up to a constant multiple)
action of ${\rm TFS}^2$ 
carrying $\Tw$ to $\YZb$
must be.  In particular, the single action of the Einstein tensor is
\beq\label{eloop4}
\begin{array}{rl}
(\underline{\underline{{\bf b}}}(b)\F)_{\l\a\b}&=
-\dfrac{n^2-2n-2}{2(n+2)(n-2)}\left\{\F_\a b_{\b\l}-\F_\b b_{\a\l}\right\} \\
&\quad +\dfrac1{2(n+2)}\Big\{\g_{\m\l}(\F_\a b_\b{}^\m-\F_\b b_\a{}^\m)\\
&\quad -g_{\b\l}\F_\m b_\a{}^\m+g_{\a\l}\F_\m b_\b{}^\m\Big\}\\
&\quad -\dfrac{n}{2(n+2)(n-2)}\left\{\g_{\b\m}\F_\a b^\m{}_\l
                            -\g_{\a\m}\F_\b b^\m{}_\l\right\} \\
&\quad +\dfrac1{2(n+2)(n-2)}\Big\{\g_{\b\l}\F_\m b_\a{}^\m-\g_{\a\l}\F_\m 
b_\b{}^\m\\
&\quad +g_{\a\l}\g_{\b\m}\F_\nu b^{\m\nu}
                           -g_{\b\l}\g_{\a\m}\F_\nu b^{\m\nu}\Big\} \\
&\quad -\dfrac1{(n+2)(n-2)}\left\{\g_{\a\b}\F_\m b^\m{}_\l
                          +\g_{\b\m}\F_\l b_\a{}^\m
                          -\g_{\a\m}\F_\l b_\b{}^\m\right\}.
\end{array}
\eeq

Direct computation, now keeping track of curvature terms, shows that

\begin{proposition} A realization of the relation \nnn{loopprinc}
corresponding to loop $\fbox{{\rm 4}}$ is given by
$$
\gZYZ\;\gZ-\gYYZ\;\gY=\dfrac2{n-2}\;
\underline{\underline{{\bf b}}}(b)
+ \underline{\underline{{\bf C}}}(C),
$$
where the action $\underline{\underline{{\bf b}}}(b)$ 
of the Einstein tensor is given by
\nnn{eloop4}, and the action of the Weyl tensor is
\bdm
\begin{array}{rl}
\Big(\underline{\underline{{\bf C}}}(C)\F \Big)_{\l\a\b}&:=
\dfrac16\;\g_{\m\nu}\left(-\F_\l 
C_{\a\b}{}^{\m\nu}
+\F_{[\a}C_{\b]\l}{}^{\m\nu}\right) \\
&\quad +\dfrac1{3n(n+2)(n-2)}\;\left(\g_{\m\l}\F_\nu C_{\a\b}{}^{\m\nu}
-\g_{\m[\a}C_{\b]\l}{}^{\m\nu}\F_\nu\right) \\
&\quad -\dfrac{2n^3-4n^2-7n+8}{2n(n+2)(n-2)}\;\F_\m C_{\a\b}{}^\m{}_\l\\
&\quad +\dfrac{2n^2-9n-6}{6n(n+2)(n-2)}\;\g_{\m\l}\F_\nu C_{\a\b}{}^{\nu\m} \\
&\quad +\dfrac{4n^2-6n-15}{3n(n+2)(n-2)}\;\F_\nu
\g_{\m[\a}C_{\b]}{}^\m{}_\l{}^\nu\\
&\quad +\dfrac{(n+3)(2n-3)}{3n(n+2)(n-2)}\;\F_\nu
\g_{\m[\b}C_{\a]}{}^{\nu\m}{}_\l \\
&\quad +\dfrac{5n+8}{6n(n+2)(n-2)}\;\left(\g_{\m\l\nu[\b}C_{\a]}{}^{\r\m\nu}
+\g_{\a\b\m\nu}C_\l{}^{\r\m\nu}\right)\F_\r \\
&\quad +\dfrac{n^2-n-8}{2n(n+2)(n-2)}\;\g_{\m\nu} \F_\r 
g_{\l[\b}C_{\a]}{}^{\r\m\nu}.
\end{array}
\edm
Here square brackets denote antisymmetrization, and 
we have employed the fourth-degree antisymmetric Clifford symbols
$$
\g_{\a\b\l\m}=\g_{[\a}\g_\b\g_\l\g_{\m]}\,.
$$
\end{proposition}

\section{Overdetermined systems}\setcounter{equation}0
Consider the following systems of differential equations on twistors
$\f$:
\beq\label{over2}
\GT\f=\t^2\f,\qquad\GY\f=0,\qquad\GZ\f=0,
\eeq
\beq\label{over0}
\GT\f=\t^2\f,\qquad\GS\f=0,\qquad\GZ\f=0,
\eeq
where $\t$ is a fixed but arbitrary smooth real function.
By virtue of Lemma \ref{strongell},
each of these systems is overdetermined.  However, each has nontrivial
solutions on the sphere $S^n$, as we shall show presently.
It is thus reasonable to ask whether a given one of these systems
{\em characterizes} the sphere, in the sense that no other manifold
supports solutions. 
If \nnn{over2} or \nnn{over0} fails to characterize the sphere, one might
ask for a classification of the manifolds that are capable of supporting
a solution.

\begin{lemma}\label{EliGS} Let $n\ge 4$.  If
\nnn{over2} holds for 
some smooth real function $\t$, then the pointwise equation
\beq\label{EliGSeqn}
\begin{array}{l}
\tfrac12\left\{\left(n-\frac74\right)^2+\frac{15}{16}\right\}
C\diamond\f+\dfrac{(n-3)n}{n-2}\;b\cdot\f \\
\qquad+(n-3)\left\{n^2\t^2
-\dfrac{(n-2)(n+2)}{4(n-1)}K\right\}\f=0
\end{array}
\eeq
holds.  In particular, if 
\beq\label{EliGShypS}
\cS_{{\bf T}} \f=\t\f\qquad\GY\f=0,\qquad\GZ\f=0
\eeq
then \nnn{EliGSeqn} holds.
On a given manifold, 
the system \nnn{over2} with $\t$ constant can have a nonzero solution
$\f$ only for finitely many values of $\t$.
\end{lemma}

\begin{proof}{Proof} The identity \nnn{EliGSeqn} is an immediate
consequence of \nnn{elimGS}.  It shows that the possible values of
$\t^2$ are bounded by a constant times 
$\max_x\|\Rop_x\|$.  But since $\GT$ is strongly elliptic
(Lemma \ref{strongell}), its
eigenvalues $\t_0^2\le\t_1^2\le\cdots$
have Weyl asymptotics $\t_j^2\sim{\rm const}\cdot j^{2/n}$
as $j\to\infty$.  Thus only finitely many $\t_j^2$
can satisfy the curvature operator bound.
\end{proof}

\section{Spectra on the sphere}\setcounter{equation}0
Despite being badly overdetermined, the system
\nnn{over2} 
does have solutions, for a certain constant $\t^2$, on the sphere $S^n$.  
By the branching rule and Frobenius reciprocity, the 
$\spN$-types of sections of $\Tw$ for $n\ge 5$ odd have highest weight labels
$$
\a_{j,k,\pm}=
\left(\frac32+j,\frac12+k,\frac12,\cdots,\frac12,\pm\frac12\right),
$$
where $j$ runs over the natural numbers, and $k$ runs over $\{0,1\}$.
Each type occurs with multiplicity one.
By \cite{tbsw}, Theorem 4.1, 
$$
\cN(\GZ)=\bop_{j=0}\a_{j,k,\pm}\,\qquad
\cN(\GY)=\bop_{k=0}\a_{j,k,\pm}\,
$$
where we have abused notation slightly by writing 
the highest weight to represent the $\spN$-type
which it labels.  Each $\a_{j,k,\pm}$ consists of eigensections
of $\cS_{{\bf T}}$, and thus of $\GT$, since $\cS_{{\bf T}}$ is 
$\spin(n+1)$-invariant:
by Schur's Lemma and the fact the $\spin(n+1)$-types occur with
multiplicity one, 
$\cS_{{\bf T}}$ must act on each $\spin(n+1)$ as multiplication by a constant.
Thus the $\spN$-types $(\frac32,\frac12,\cdots,\frac12,\pm\frac12)$
consist of solutions of \nnn{over2}, and choosing just one of these
two types, one gets solutions of \nnn{EliGShypS}.
When $n$ is even, the section space 
of each (positive and negative) twistor
bundle is a (multiplicity one) direct sum of $\spN$-modules with
labels
$$
\a_{j,k}=
\left(\frac32+j,\frac12+k,\frac12,\cdots,\frac12\right),
$$
where $j$ runs over the natural numbers, and $k$ runs over $\{0,1\}$.
Again, the solutions of \nnn{over2} are the summands with $j=k=0$.

The spectra on the sphere 
of all the operators we study here, and in fact of 
any operator of the form $G_{\l\s_u}^*G_{\l\s_u}$ 
on any irreducible $\spin(n)$-bundle
$\bbV(\l)$,
are given in \cite{tbsw}, Theorems 4.1 and 5.1.  
(The first of these theorems 
gives the spectrum to within an overall normalizing
constant, and the second computes the normalizing constant.)
The branching rule and Frobenius
reciprocity show that the $\spin(n+1)$ types $\a$ occurring in the space
of sections of $\bbV(\l)$ occur with multiplicity one, and 
are exactly those satisfying the {\em interlacing rule}
\beq\label{interlace}
\begin{array}{ll}
\a_1\ge\l_1\ge\cdots\ge\l_\h\ge|\a_{\h+1}|,\qquad & n\ {\rm odd}, \\
\a_1\ge\l_1\ge\cdots\ge\a_\h\ge|\l_\h|,\qquad & n\ {\rm even}.
\end{array}
\eeq
For the operator $G_{\l\s_u}^*G_{\l\s_u}\,$, the eigenvalue
on the $\a$ summand is
\begin{equation}\label{theeig}
c_{\l\s_u}\prod_{a=1}^L(\tilde\a_a^2-s_u^2)=\tilde
c_{\l\s_u}\prod_{a\in\cT(\l)}(\tilde\a_a^2-s_u^2),
\end{equation}
where $L=[(n+1)/2]$, $s_u$ is the quantity defined in 
(\ref{casval},\ref{defsu}),
$$
\tilde\a_a=\a_a+\dfrac{n+1-2a}2\,,
$$
$\cT(\l)$ is the set of all $a$ in $\{1,\ldots,L\}$ for which 
$\tilde\a_a^2$ is allowed only one value by the interlacing
rule \nnn{interlace}, and $c_{\l\s_u}\,$, $\tilde c_{\l\s_u}$
are certain normalizing constants.  Given $\l$ and $\tilde c_{\l\s_u}\,$,
one may compute $c_{\l\s_u}\,$, so it suffices to describe
$\tilde c_{\l\s_u}\,$.  Let $t(\l)$ be the cardinality of $\cT(\l)$.
By \cite{tbsw}, Theorem 5.2,
\bdm
\tilde c_{\l\s_u}=\left\{\begin{array}{ll}
\dfrac{(-1)^{t(\l)+1}}{\prod_{1\le v\le N(\l),\;v\ne u}(s_v-s_u)}\,,\qquad
& N(\l)\ {\rm odd}, \\\
\dfrac{(-1)^{t(\l)+1}}{2\prod_{1\le u\le N(\l),\;(\s_u)_\h=0}(s_u+\frac12)}\,,
\qquad & n\ {\rm even},\ \l_\h=0\ne\l_{\h-1},\ |(\s_u)_\h|=1, \\\
\dfrac{(-1)^{t(\l)}(s_u+\frac12)}{\prod_{1\le v\le N(\l),\;v\ne u}(s_v-s_u)}
\qquad
& {\rm otherwise}.
\end{array}\right.
\edm
(A unified formula handling all cases is given in \cite{tbsw}, Remark 5.6.)

In the present situation, let 
$$
\begin{array}{llll}
&s_{{\bf\S}}=\dfrac{n}2\,,\ \ &\tc_{{\bf\S}}=-\dfrac1{n(n-1)},\\ 
&s_{{\bf T}}=0\,,\ \ &\tc_{{\bf T}}=\dfrac4{n(n+2)(n-2)}\,,\\ 
&s_{{\bf Y}}=-\dfrac{n-2}2\,,\ \ &\tc_{{\bf Y}}=\dfrac{n-3}{2(n-1)(n-2)}\,, \\
&s_{{\bf Z}}=-\dfrac{n+2}2\,\ \ &\tc_{{\bf Z}}=-\dfrac1{2(n+2)}\,.
\end{array}
$$
In odd dimensions, these describe the above constants corresponding
to the gradient targets $\Sp$, $\Tw$, $\Yb$, and $\Zb$ in that order.
In even dimensions, starting with the bundle 
$$
\Tw^{\pm}\cong_{\spin(n)}\bbV\left(\frac32\,,\frac12\,,\ldots,
\frac12\,,\pm\frac12\right),
$$
we get the constants corresponding to the targets $\Sp^\pm\,$,
$\Tw^{\mp}\,$, $\Yb^{\pm}\,$, and $\Zb^{\pm}$ in that order.
As a consequence, we have the following spectra.  On the 
$\spin(n+1)$-type 
$$
\a(j,k,\e):=\left(\frac32+j,\frac12+k,\frac12,\ldots,\frac12\,,
\e\frac12\right),
$$
where $j\in\bbN$, $k\in\{0,1\}$, and 
\bdm
\e\left\{\begin{array}{ll}
\in\{-1,1\},\ & n\ {\rm odd}, \\
=1, & n\ {\rm even},
\end{array}\right.
\edm
the eigenvalues are:
$$
\begin{array}{cc}
\underline{{\rm operator}} & \underline{{\rm eigenvalue}} \\
\GS\qquad & \dfrac{(j+n+1)(j+1)(1-k)}{n} \\
\GT=(\cS_{{\bf T}})^2\qquad & \dfrac4{n(n+2)(n-2)}\left(j+\dfrac{n}2+1\right)^2
\left(k+\dfrac{n}2-1\right)^2 \\
\GY & \dfrac{(n-3)(j+n)(j+2)k}{2(n-2)} \\
\GZ & \dfrac{j(j+n+2)\{2n-(n-1)k\}}{2(n+2)}
\end{array}
$$
Note that \nnn{theeig} gives formulas for the eigenvalues that
are quadratic in $k$; but since $k^2=k$, we may reduce them to
linear-in-$k$ expressions if we like.
Note also that when
$n$ is even, the section space of $\Tw^+$ contains one copy of
the $\spin(n+1)$-type $\a(j,k,1)$, and the 
section space of $\Tw^-$ also contains one copy.

The leading asymptotics in $j$ as $j\to\infty$ in the above list
coincide with the arrays \nnn{block} and \nnn{blocks}, of eigenvalues
of the leading symbols on arbitrary manifolds.  This is an example
of a more general phenomenon explored in detail in \cite{ab2}.

In particular, the above eigenvalue list shows that
\beq\label{annih2}
\begin{array}{l}
\GS\ {\rm annihilates\ the}\ k=1\ {\rm summands}, \\
\GY\ {\rm annihilates\ the}\ k=0\ {\rm summands}, \\
\GZ\ {\rm annihilates\ the}\ j=0\ {\rm summands}.
\end{array}
\eeq
In fact, \nnn{annih2} is predictable from the fact that 
the sections spaces of $\Sp$, $\Yb$, $\Zb$ do not contain
copies of $\a(j,1,\e)$, $\a(j,0,\e)$, $\a(0,k,\e)$ respectively;
the operators $\gS$, $\gY$, $\gZ$, which are targeted in these bundles,
must already annihilate the relevant summands.

The summands $\a(0,0,\e)$ in the $\spin(n+1)$-decomposition of the
$\spin(n+1)$-finite section space thus satisfy the system \nnn{over2},
with 
$$
\t^2=\dfrac4{n(n+2)(n-2)}\left(\dfrac{n}2+1\right)^2
\left(\dfrac{n}2-1\right)^2.
$$
Similarly, the $\a(0,1,\e)$ summands satisfy the system \nnn{over0},
with 
$$
\t^2=\dfrac4{n(n+2)(n-2)}\left(\dfrac{n}2+1\right)^2
\left(\dfrac{n}2\right)^2.
$$

\section{Interaction with sharp Kato estimates}\label{wkato}
\setcounter{equation}0

Suppose we have a natural irreducible bundle $\bbV(\l)$, with
gradients
$$
G_u:\bbV(\l)\to\bbV(\s_u),\qquad u=1,\ldots,N(\l).
$$
Partition $\{1,\ldots,N(\l)\}$ into two sets $A$ and $A^{{\rm c}}$, and
suppose that $\f$ is a smooth section of $\bbV(\l)$ on a compact
manifold $M$ satisfying
\beq\label{Aeqns}
G_u\f=0,\qquad{\rm all}\ u\in A.
\eeq
(We assume that $M$ has whatever structure necessary to support
the bundles -- SO$(n)$ or Spin$(n)$.)
Choose a 
BW formula
$$
\sum_{u=1}^{N(\l)}t_uG_u^*G_u={\rm Curv},
$$
where ``Curv'' is an action of the Riemann curvature on $\bbV(\l)$.
Then
\beq\label{startkato}
\begin{array}{rl}
\int_M({\rm Curv}\,\f,\f)&=\int_M\sum_{u\in A^{{\rm c}}}t_u(G_u^*G_u\f,\f) \\
&=\int_M\sum_{u\in A^{{\rm c}}}t_u|G_u\f|^2 \\
&\ge(\min_{u\in A^{{\rm c}}}t_u)\int_M
\underbrace{\sum_{u=1}^{N(\l)}|G_u\f|^2}_{|\N\f|^2}.
\end{array}
\eeq

Associated to the set $A$ is a {\em sharp Kato constant} $k_A$
\cite{tbkato,cgh}.  This is the best universal constant 
with the (local) property
$$
G_u\f=0,\ {\rm all}\ u\in A\ \Rightarrow\ 
\bigg|d|\f|\bigg|^2\le k_A|\N\f|^2\ {\rm off}\ \{\f_x=0\}.
$$
If the system \nnn{Aeqns} is injectively elliptic, then $k_A<1$;
otherwise $k_A=1$.  In fact, $k_A$ is $1-\e_A$, where $\e_A$
is the best ellipticity constant for $\sum_{u\in A}G_u^*G_u$:
$$
\s_2\left(\sum_{u\in A}G_u^*G_u\right)(\x)\ge\e_A|\x|^2.
$$
(See \cite{tbkato}, Theorems 4 and 7.)
Ellipticity constants like this can be computed from arrays like
\nnn{block} and \nnn{blocks}.
This and \nnn{startkato} give
\beq\label{ctkato}
\int_M({\rm Curv}\,\f,\f)\ge\dfrac{m_A}{k_A}\int_M\bigg|d|\f|\bigg|^2,
\eeq
where
$$
m_A=\min_{u\in A^{{\rm c}}}t_u.
$$
By a standard argument, the restriction ``off $\{\f_x=0\}$'' 
disappears upon integration; this depends on the fact
that for any smooth section $\f$, the scalar quantity 
$|\f|$ is a distribution in 
the Sobolev space $L^2_1$.

Now assume all integrals are taken with respect to {\em normalized}
measure.  (The foregoing statements about integrals are insensitive
to normalization of the measure.)
Consider the orthogonal (Hodge) decomposition of $|\f|$ as
$$
|\f|=|\f|_{{\rm const}}+|\f|_{{\rm div}},
$$
where $|\f|_{{\rm const}}$ is a constant function, and 
$|\f|_{{\rm div}}$ is in the $L^2$-span of the eigenfunctions
of the scalar Laplacian $\D$ which have positive eigenvalues.
If $0=\m_0<\m_1\le\cdots$ are the eigenvalues of $\D$, with
corresponding orthonormal eigenfunctions $\y_j$, and 
$$
|\f|=\sum_{j=0}^\infty a_j\y_j,
$$
then $\y_0=1$ and 
$$
|\f|_{{\rm const}}=\int|\f|=\|f\|_1
$$
is the $L^1$ norm of $\f$ in normalized measure.
Integrating by parts on the right in \nnn{ctkato}, we have
$$
\begin{array}{rl}
\int_M({\rm Curv}\,\f,\f)&\ge\dfrac{m_A}{k_A}\int_M
|\f|_{{\rm div}}\D|\f|_{{\rm div}} \\
&\ge\dfrac{\m_1m_A}{k_A}\int_M|\f|_{{\rm div}}^2 \\
&=\dfrac{\m_1m_A}{k_A}\bigg\||\f|_{{\rm div}}\bigg\|^2_2.
\end{array}
$$
As a result, if we have an assumption
$$
{\rm Curv}\le C{\rm Id}_{\bbV(\l)}
$$ 
in the sense of endomorphisms, then
we may conclude that
\bdm
\bigg\||\f|_{{\rm div}}\bigg\|^2_2\le\dfrac{k_AC}{\m_1m_A}\|\f\|_2^2.
\edm
Note that this estimate is scale invariant (as it must be):
if the metric $g$ is rescaled to $A^2g$, where $A$ is a positive
constant, then both $C$ and $\m_1$ scale by factors of $A^{-2}$,
while the other constants remain fixed.  Thus the part
of $|\f|$ which is orthogonal to the constants cannot be too
large; in this sense, $|\f|$ is approximately constant.

Since 
$$
\bigg\||\f|_{{\rm div}}\bigg\|^2_2=
\bigg\|\f\bigg\|^2_2-\bigg\||\f|_{{\rm const}}\bigg\|^2_2,
$$
we could also write this as
$$
\bigg\||\f|_{{\rm const}}\bigg\|^2_2\ge\left(1-\dfrac{k_AC}{\m_1m_A}\right)
\|\f\|^2_2.
$$
Since $|\f_{{\rm const}}|=\|\f\|_1$,
this relates the $L^2$ and $L^1$ norms of $\f$:
$$
\|\f\|_1^2\ge\left(1-\dfrac{k_AC}{\m_1m_A}\right)
\|\f\|^2_2.
$$
In general, for any section, since the measure has been normalized,
$\|\f\|_1\le\|\f\|_2$ (by the convexity of $x\mapsto x^2$), so 
we have
\beq\label{LoneLtwo}
\left(1-\dfrac{k_AC}{\m_1m_A}\right)
\|\f\|^2_2\le\|\f\|_1^2\le\|\f\|_2^2.
\eeq

This is a statment of the approximate constancy of $|\f|$, which
becomes stronger as $C$ gets smaller.  (If $C\le 0$, $|\f|$ must 
be constant; in fact \nnn{startkato} implies the stronger statement
that $\f$ is parallel.)
The use of the improved Kato inequality (resulting in the
appearance above of $k_A$ in place
of $1$) also improves things somewhat.
If we have a choice of BW formulas (i.e., if $N(\l)\ge 4$), then
the interplay between $m$ and $C$ can be somewhat complicated.

If there is a self-gradient and its index $u_0$ lies in $A$, we may
replace the condition $G_{u_0}\f=0$ with the eigensection equation 
$D_{{\rm self}}\f=\eta\f$ (see below for definitions)
without disturbing the improved Kato 
inequality.
Recall that for a self-gradient, we need a summand of 
$T^*M\otimes\bbV(\l)$ to be isomorphic to $\bbV(\l)$ itself; that is, 
there must be a bundle map in
$$
0\ne \z\in\hom_H(\bbV,T^*M\otimes\bbV).
$$ 
This results in a natural first-order differential operator on
the original realization of $\bbV$, namely
\beq\label{selfgrad}
D_{{\rm self}}:=-\z^*\circ\N=-\z^*\circ{\rm Proj}_{[\bbV]}\circ\N=-\z^*\circ G_s\,;
\eeq
this is the {\em self-gradient}.
Since the difference between two Spin$(n)$-connections on $\bbV$ is an element
of $\hom_{{\rm Spin}(n)}
(\bbV,T^*M\otimes\bbV)$, we also have the family of modified
$H$-connections
\bdm
\tN:=\N+a\z,\ \ a\in\bbR.
\edm
In the spinor case, these are known as {\em Friedrich connections}.
Using $\tN$ instead of $\N$ in the formula \nnn{selfgrad} for the
self-gradient results in 
$$
\tilde D=D-a\z^*\z.
$$

Though $\bbV$ need not have a distinguished real form,
there is a distinguished real form of $\bbV\ot\bbV^*\cong\endo(V)$,
namely the self-adjoint endomorphisms.
Since $T^*M$ has a distinguished real form, it makes sense to demand
that $\z$ be {\em imaginary}; this is what is required to make
the self-gradient $D$ formally self-adjoint.
For example, in the spinor case, we construct the Clifford multiplication
$\g$ so that each $\g(\x)$ is skew-adjoint, with the result that 
the Dirac operator is formally self-adjoint.  This fixes the 
normalization of $\z$ up to a constant factor in $\bbR^*$.
If $G_s$ is the gradient valued in the summand of $T^*M\otimes\bbV$
which is $H$-isomorphic to $\bbV$, then the requirement that
\beq\label{d2}
D^2=G^*_sG_s
\eeq
fixes the normalization of $\z$ up to a factor of $\pm 1$.
This is in fact the best one can do; see \cite{tbjlt}.  For example,
replacing 
$\g$ by $-\g$ has no effect on spinor theory.  Ultimately, all these
ambiguities are rooted in the fact that $\eye\mapsto-\eye$ is an
automorphism of $\bbC$.

By \nnn{d2} and \nnn{selfgrad},
$$
G_s^*\z\z^*G_s=G_s^*G_s\,,
$$
so $\z\z^*$, being an $H$-map on $T^*M\ot\bbV$, is the 
projection on the $\bbV(\l)$-isomorphic summand, and $\z^*\z$ is the identity
on $\bbV$.

Choosing $\z$ to be imaginary also has the effect of making the 
natural metrics on $\bbV(\l)$ and $T^*M\otimes\bbV(\l)$ {\em compatible}
with the modified connections $\tN$; that is, $\tN h=0$ 
whenever $h$ is one of these metrics.  As noted in \cite{cgh},
the sharp Kato constants remain unchanged upon 
passage from $\N$ to a new compatible connection.
(One can also easily observe this by
examining the argument of
\cite{tbkato}.)

In the case of twistors, the resulting statements build on, for example, 
Theorems \ref{van} and \ref{bbounds}, giving a weaker conclusion under
a relaxed assumption.  
There are clearly many results along these lines that could be stated;
lacking an immediate application, 
we shall content ourselves here with just a few.
Using data from \cite{tbkato} or computing directly from
\nnn{block} and \nnn{blocks} to get Kato constants, we have
the following relatives of Theorem \ref{van} and 
Proposition \ref{dim8}:

\begin{theorem}\label{usekato}
Let $Q$ be the maximum eigenvalue over $M$ of the curvature operator 
on $\L^2$.
With normalized measure,
for a section $\f$ of $\bT$ with $\gZ\f=0$,
$$
\left(1-\dfrac{n(n+1)(n+7)Q}{8(n+2)\m_1}\right)\|\f\|^2_2\le\fTH\le\fHH.
$$
\end{theorem}

\begin{proof}{Proof} This is the basic estimate \nnn{LoneLtwo}, 
based on the BW formula of \nnn{gauresult}.
by \nnn{GauTwi}, $m_A=1/2$, and by \nnn{blocks}, 
$k_A=(n+1)/(2(n+2))$.  By \nnn{GauTwi} and \nnn{RopandRq}, the
quantity $C$ in \nnn{LoneLtwo} may be taken to be $n(n+7)Q/8$.
\end{proof}

\begin{theorem}\label{dimeight}
Let $n=8$, and let $\underline{B}$, $\overline{B}$ be the minimum and
maximum eigenvalues of the trace-free Ricci tensor $b$ over $M$.
For an eigensection of the Rarita-Schwinger operator with $\gZ\f=0$,
$$
\left(1-\dfrac{7\overline{B}}{18\m_1}\right)\|\f\|^2_2\le\fTH\le\fHH.
$$
For an eigensection of the Rarita-Schwinger 
operator with $\gS\f=0$, $\gY\f=0$, 
$$
\left(1+\dfrac{2\underline{B}}{5\m_1}\right)\|\f\|^2_2\le\fTH\le\fHH.
$$
\end{theorem}

\begin{proof}{Proof} Both assertions are based on the BW formula
\nnn{elimck}.  For the first statement, we may 
take $m_A=3$, $C=2\overline{B}$ (by \nnn{bac}),
and $k_A=7/12$ (by \nnn{block} and \nnn{blocks}).  For the second
statement, we may take $m_A=1$, $C=-2\underline{B}$, and
$k_A=1/5$.
\end{proof}

\small\renewcommand{\section}{\subsubsection}

\ 

Thomas Branson\newline
Department of Mathematics, The University of Iowa, Iowa City
IA 52242 USA\newline
branson@math.uiowa.edu

\

Oussama Hijazi\newline
Institut Elie Cartan, Universite Henri Poincare, Nancy 1,  
B.P.\ 239, 54506 Vandoeuvre-L{\`e}s-Nancy Cedex, France\newline
hijazi@iecn.u-nancy.fr

\end{document}